\documentclass[10pt, a4paper]{scrartcl}

\usepackage{latexsym}
\usepackage{amssymb}
\usepackage[nonamelimits]{amsmath}
\usepackage[authoryear, round]{natbib}
\usepackage{url}

\usepackage{color}
\definecolor{myred}{rgb}{0.5,0,0}
\definecolor{myblue}{rgb}{0,0,0.75}
\definecolor{mygreen}{rgb}{0,0.5,0}

\usepackage{ifpdf}

\ifpdf
   \usepackage[pdftex]{graphicx}
   \usepackage[pdftex,
               pdftitle={},
               pdfsubject={},
               pdfkeywords={},
               pdfauthor={Dirk Tasche},
               pdfstartview=FitR,
               breaklinks=true,
               colorlinks=true,
               citecolor=myred,
               linkcolor=myblue,
               urlcolor=mygreen]{hyperref}
\else
   \usepackage[dvips]{graphicx}
   \usepackage{hyperref}
   \usepackage{rotating}
\fi

\newtheorem{theorem}{Theorem}[section]

\newtheorem{remark}[theorem]{Remark}
\newtheorem{example}[theorem]{Example}
\newtheorem{proposition}[theorem]{Proposition}
\newtheorem{definition}[theorem]{Definition}

\newcommand{\qed}{$\Box$}

\numberwithin{equation}{section}

\title{Fitting a distribution to Value-at-Risk and Expected Shortfall, 
with an application to covered bonds}

\author{%
Dirk Tasche\thanks{E-mail: dirk.tasche@gmx.net\newline
The author currently works at the Prudential Regulation Authority (a
division of the Bank of England). He is also a visiting professor at
Imperial College, London. 
The opinions expressed in this paper are those of the author 
and do not necessarily reflect views of the Bank of England.}}

\date{}

\begin{document}

\maketitle

\begin{abstract}
Covered bonds are a specific example of senior secured debt. If the issuer of the bonds defaults the proceeds of the assets in the cover pool are used for their debt service. If in this situation the cover pool proceeds do not suffice for the debt service, the creditors of the bonds have recourse to the issuer's assets and their claims are pari passu with the claims of the creditors of senior unsecured debt.  Historically, covered bonds have been very safe investments. During their more than two hundred years of existence, investors never suffered losses due to missed payments from covered bonds. From a risk management perspective, therefore modelling covered bonds losses is mainly of interest for estimating the impact that the asset encumbrance by the cover pool has on the loss characteristics of the issuer's senior unsecured debt.  We explore one-period structural modelling approaches for covered bonds and senior unsecured debt losses with one and two asset value variables respectively. Obviously, two-assets models with separate values of the cover pool and the issuer's remaining portfolio allow for more realistic modelling. However, we demonstrate that exact calibration of such models may be impossible. We also investigate a one-asset model in which the riskiness of the cover pool is reflected by a risk-based adjustment of the encumbrance ratio of the issuer's assets.
\\[1ex]
\end{abstract}

\section{Introduction}
\label{se:intro}

Covered bonds are special debt instruments\footnote{%
The European Covered Bond Council on their website \href{http://ecbc.hypo.org/}{http://ecbc.hypo.org/} 
provides detailed information about the definitions and
properties related to covered bonds.}  that are normally served by their issuer. But if the issuer 
defaults and is no longer able to make the payments due to the covered bonds holders, the debt service is 
continued on the basis of the proceeds of the assets in the pool covering the bonds. If in the case of
the bond issuer's default the proceeds of the cover pool assets do not suffice for the debt service and the 
bond holders claims are not met by the liquidation of the assets, the bond holders still can take recourse
to the remaining assets of the issuer. Their claims are then pari passu with the claims of the holders of
the issuer's senior unsecured debt. 

Thus the covered bonds holders have a privileged position compared to 
most other of the issuer's creditors. Indeed, no loss due to missed payments was ever observed for European covered bonds \citep{EUBriefingCoveredBonds}. While this is good news for covered 
bonds investors, it makes estimation of expected loss for covered bonds difficult.

Covered bonds have been popular investments for a long time especially in Europe. 
As a consequence of the recent financial crisis, covered bonds became an even more important funding
instrument for banks, as such replacing instruments like securitised products.
Therefore, asset encumbrance in the banks' balance sheets due to the higher demand of collateral is increasing. 
This has led to concerns that expected loss for creditors of  senior unsecured debt is also increasing
\citep{CGFS49Paper}. But it is not clear how
to efficiently quantify this effect for the purpose of loss provisioning or stress testing.

This paper draws on recent work on fitting lognormal asset value distributions to given probability of
default (PD) and loss-given-default (LGD) estimates 
\citep{yang2015volatility} and on asset encumbrance \citep{chan-lau&Oura2014bail} to 
suggest simple and intuitive
models for covered bonds that allows quantitative assessment of expected loss and the impact of
asset encumbrance.

We find that one-period lognormal two-assets model in some situations cannot exactly be calibrated even if
different time horizons for bond issuer and covered bond holders are taken into account. The proposed 
adjusted lognormal one-asset model represents an acceptable workaround for the calibration issues, but only
as long as the credit quality fo the cover pool is only moderately better than the bond issuer's credit quality.

The paper is organised as follows:
\begin{itemize}
\item In Section~\ref{se:VaRES}, we revisit the PD and LGD fitting observation by \citet{yang2015volatility},
proving that there is a unique solution for the two model parameters and placing the approach in the 
general context of the method of moments.
\item In Section~\ref{se:CovBonds}, we extend the lognormal one-asset model by \citet{chan-lau&Oura2014bail} to 
a lognormal two-assets model and provide formulae for efficient numerical computation of expected losses
for covered bonds, senior unsecured debt and junior debt with this model.
\item Computations for the two-assets model are straightforward when the model parameters are known.
However, we show in Section~\ref{se:2lognormal} that the calibration of the parameters leads
to a non-linear equation system that may have no exact solution at all.
\item In Section~\ref{se:1lognormal}, we revisit the one-asset model by 
\citet{chan-lau&Oura2014bail}.
We adjust the model for portfolio heterogeneity in terms of risk by modifying the asset encumbrance ratio
related to the covered bonds to reflect the expected loss of the cover pool. We show that in the case of
a homogeneous portfolio the adjusted lognormal one-asset
model is equivalent to a comonotonic special case of the two-assets model discussed in
Section~\ref{se:CovBonds}. Unsurprisingly, therefore, exact calibration of the model again is not 
always possible.
\item In Section~\ref{se:numerical}, we provide some numerical examples illustrating the impact of 
asset dependence, of asset encumbrance, and of approximating the two-assets model with the adjusted one-asset model.
\item The paper concludes with some comments in Section~\ref{se:conclusions}.
\item Appendix~\ref{se:MeanVar} provides background information on mean-variance matching.
\end{itemize}

\section{Quantile-Expected Shortfall matching}
\label{se:VaRES}

The \emph{Method of Moments}\footnote{%
See Appendix~\ref{se:MeanVar} for a brief description of its simplest version,
the mean-variance matching.} probably is the most popular approach to fitting a distribution 
to a finite set of given 
characteristics. But it is not the only conceivable approach. 
\citet{hosking1992moments}, for instance, argued that matching with so-called 
L-moments might provide results that are 
better aligned with the results of standard goodness-of-fit tests. In a finance-related context, \citet{Tasche2009a}
suggested \emph{quasi-moment matching} for the probability-of-default curves of rating or scoring models in order to
appropriately reflect the discriminatory power of the rating models.
In a recent paper, \citet{yang2015volatility} observed that lognormal distributions may be uniquely 
determined by specified
PD (probability of default) and recovery rate (or, equivalently, LGD). 

Assume that $X$ is an integrable real-valued random variable which
is interpreted as showing the value of a financial asset at some future point in time -- say one year in the
future.
Of course, the asset value interpretation holds only if $X$ is a non-negative random variable. Nonetheless, the 
following concepts apply to any real random variable $X$:
\begin{subequations}
\begin{itemize}
\item Let $D \in \mathbb{R}$ (interpreted as debt due for repayment at the time horizon). Then the \emph{PD} (probability of default)
associated with $D$ is 
\begin{equation}\label{eq:PD}
PD \ = \ \mathrm{P}[X < D].
\end{equation}
\item  Let $D \in \mathbb{R}\backslash\{0\}$ such that $PD >0$. Then the \emph{RR} (recovery rate) associated with $D$ is\footnote{%
The \emph{indicator function} $\mathbf{1}_E$ of the event $E$ is defined by $\mathbf{1}_E(\omega) = 1$ for $\omega \in E$ and
$\mathbf{1}_E(\omega) = 0$ for $\omega \notin E$.}
\begin{equation}\label{eq:RR}
RR \ = \ \frac{\mathrm{E}[X\,|\,X < D]}{D} \ = \ \frac{\mathrm{E}[X\,\mathbf{1}_{\{X < D\}}]}{D\,\mathrm{P}[X < D]}.
\end{equation}
\end{itemize}
\end{subequations}
Stating the recovery rate is equivalent to stating the \emph{LGD} (loss given default) which is defined as
$LGD = 100\% - RR$.

\begin{remark}\label{rm:PDvQuant}
In case of $X$ having a distribution with positive density, 
by \eqref{eq:PD} the threshold value $D$ is the quantile at level
$PD$ of $X$, i.e.
\begin{equation*}
    D \ = \ q_{PD}(X),
\end{equation*}
with $q_\alpha$ defined in general by $q_\alpha(Z) = \min\{z: \mathrm{P}[Z \le z] \ge \alpha\}$ for $0 < \alpha < 1$ and
any real-valued random variable $Z$. In the finance community, quantiles associated with loss variables 
are often referred to as Value-at-Risk (VaR). 

Similarly, in case of $X$ with a distribution with positive density, the recovery rate associated with $D$ is closely related to the
\emph{expected shortfall} (ES) of $X$ at level $PD$:
\begin{equation*}
    RR \ = \ \frac{\mathrm{ES}_{PD}(X)}{D},
\end{equation*}
with $\mathrm{ES}_\alpha$ defined in general by 
\begin{equation*}
\begin{split}
\mathrm{ES}_\alpha(Z)& \ = \alpha^{-1} \int_0^\alpha q_u(Z)\,du\\
& \ = \ \mathrm{E}[Z\,|\,Z \le q_\alpha(Z)] - (q_\alpha(Z) - \mathrm{E}[Z\,|\,Z \le q_\alpha(Z)])
    \left(\frac{\mathrm{P}[Z \le q_\alpha(Z)]}\alpha - 1\right).
\end{split}
\end{equation*} 
for $0 < \alpha < 1$ and
any real-valued random variable $Z$. If $\mathrm{P}[Z \le q_\alpha(Z)] = \alpha$, it follows that 
$$\mathrm{ES}_\alpha(Z) \ = \ \mathrm{E}[Z\,|\,Z \le q_\alpha(Z)],$$
while in general we only have $\mathrm{ES}_\alpha(Z) \le \mathrm{E}[Z\,|\,Z \le q_\alpha(Z)]$.
In the finance community, ES often is associated with loss variables and therefore defined
as conditional expectation with respect to the upper (right-hand side) tail of the distribution
\citep{Acerbi&Tasche}.
In the literature, ES is referred to also as Conditional VaR \citep{Rockafellar&Uryasev} or Average VaR 
\citep{follmer2011stochastic}.
\end{remark}
In the following, we will study the problem of how to fit a distribution to given PD and RR (or equivalently 
to given quantile and ES). Hence, we will have two equations to determine the distribution. That is why 
is is natural to consider two parameter location-scale distribution families which are defined as follows.
\begin{definition} \label{de:LocScale}
Let $X$ be a real-valued random variable with given distribution. Then the set of the distributions of the random variables 
$m + s\,X$, $m\in\mathbb{R}$, $s > 0$ is called the \emph{location-scale distribution family} associated 
with the \emph{generating variable} $X$. 
\end{definition}
By Remark~\ref{rm:PDvQuant}, all observations we make in the following for quantile-ES matching of 
two-parameter 
distribution families immediately apply also to PD-LGD matching.

\subsection{Quantile-ES matching with location-scale distribution families}
\label{se:ESLocSc}

Assume that values $0<\alpha <1$ for the quantile confidence level, $q \in \mathbb{R}$ for the quantile,
$t < q$ for the ES as well as a location-scale distribution family with generating random variable $X$ are given.
In addition, we assume that $X$ is integrable such that $\mathrm{ES}_\alpha(X)$ is well-defined and finite.
We also assume that $\mathrm{ES}_\alpha(X) < q_\alpha(X)$ holds.

If $Y = m + s\,X$ with $m\in\mathbb{R}$ and $s > 0$ is a representation of any element of the location-scale family
then its $\alpha$-quantile and $\mathrm{ES}_\alpha$ respectively are given by
\begin{subequations}
\begin{equation}
\begin{split}
    q_\alpha(Y) & \ = \ m + s\,q_\alpha(X),\\
    ES_\alpha(Y) & \ = \ m + s\,ES_\alpha(X).
\end{split}
\end{equation}
Hence, solving $q = q_\alpha(Y)$ and $t = ES_\alpha(Y)$ for $m$ and $s$ gives
\begin{equation}\label{eq:ESSolve1}
\begin{split}
    s & \ = \ \frac{q-t}{q_\alpha(X)-ES_\alpha(X)},\\
    m & \ = \ q - \frac{q-t}{q_\alpha(X)-ES_\alpha(X)}\,q_\alpha(X).
\end{split}
\end{equation}
\end{subequations}
In the special case where $X$ is standard normal with quantile\footnote{%
$\Phi$ denotes the standard normal distribution function.}  $\Phi^{-1}(\alpha)= q_\alpha(X)$, 
we have\footnote{%
$\varphi$ denotes the standard normal density.} 
$$ES_\alpha(X)\ =\ - \frac{\varphi(\Phi^{-1}(\alpha))}{\alpha}\ <\ \Phi^{-1}(\alpha).$$ 
On the one hand,
this provides a proof of the well-known inequality 
\begin{equation}\label{eq:ineq}
	\Phi(- a) \ <\ \frac{\varphi(a)}{a},\quad\text{for}\ a > 0. 
\end{equation}
On the other hand, it follows that in the standard normal case \eqref{eq:ESSolve1} reads as follows:
\begin{equation}\label{eq:ESSolve}
\begin{split}
    s & \ = \ \frac{\alpha\,(q-t)}{\alpha\,\Phi^{-1}(\alpha)+\varphi(\Phi^{-1}(\alpha))},\\
    m & \ = \ \frac{\alpha\,t\,\Phi^{-1}(\alpha)+q\,\varphi(\Phi^{-1}(\alpha))}{\alpha\,\Phi^{-1}(\alpha)+\varphi(\Phi^{-1}(\alpha))}.
\end{split}
\end{equation}
Since the location parameter of a location-scale distribution family can be any real number, 
negative numbers may
be included in the support sets of some or all distributions in the family. Therefore, in general 
location-scale distribution families cannot be used for fitting distributions on the 
positive real axis or the unit interval.

\subsection{Quantile-ES matching with lognormal distributions}
\label{se:ESlognormal}

A simple way to generate two-parameter families of distributions
on the positive real half-axis is to take the exponential of a location-scale family 
in the sense of Definition~\ref{de:LocScale}.
Hence we study distributions of the shape 
\begin{equation}\label{eq:ExpLocSc}
Y \ = \ \exp(m + s\,X),
\end{equation}
where $X$ denotes the random variable generating the location-scale family and we assume $m \in \mathbb{R}$ 
and $s > 0$.

We want to fit a lognormally distributed random variable $Y$, as given by \eqref{eq:ExpLocSc} with $X$
standard normal, to given values of $0<\alpha<1$ for the confidence level, $q$ for the $\alpha$-quantile and
$t < q$ for the $\mathrm{ES}_\alpha$. Since $Y$ is positive we must also require $0 < t < q$.
We then have to solve the following non-linear equation system for $m$ and $s$:
\begin{subequations}
\begin{equation}\label{eq:SysLognormal}
\begin{split}
    q & \ = \ q_\alpha\bigl(\exp(m+s\,X)\bigr) \ = \ \exp(m + s\,\Phi^{-1}(\alpha)), \\
    t & \ = \ ES_\alpha\bigl(\exp(m+s\,X)\bigr) \ = \ \alpha^{-1}\,
     \exp(m + s^2/2)\,\Phi\big(\tfrac{\log(q)-m}{s}-s\bigr).   
\end{split}
\end{equation}
See, e.g., \citet[][Appendix~A]{yang2015volatility} for a derivation of the second 
row of \eqref{eq:SysLognormal}.
Rearranging the first row of \eqref{eq:SysLognormal} for $m$ and substituting the resulting term for $m$
in the second row gives
\begin{equation}\label{eq:LognormalSolve}
\begin{split}
    m & \ = \ \log(q) - s\,\Phi^{-1}(\alpha), \\
    0 & \ =\ q\,\Phi\big(\Phi^{-1}(\alpha)-s\bigr) - \alpha\,t\,\exp\bigl(s\,\Phi^{-1}(\alpha)-s^2/2\bigr).  
\end{split}
\end{equation}
\end{subequations}
\begin{proposition}\label{pr:lognormal}
For $a, b\in\mathbb{R}$ fixed define the function $f: \mathbb{R} \to \mathbb{R}$ by
$$f(s) \ = \ \Phi(a-s) - b\,\exp(a\,s-s^2/2).$$
If $0< b < \Phi(a)$ then there is exactly one $s_0 \in \mathbb{R}$ such that $f(s_0) = 0$. This number $s_0$ satisfies the
inequality
$$0\ <\ s_0\ <\ \frac{\varphi(a)}{b} + a.$$
\end{proposition}
\textbf{Proof.} Observe that $f$ has the following properties:
\begin{enumerate}
\item[(i)] $f(0) = \Phi(a) - b > 0$.
\item[(ii)]  $\lim_{s \to \infty} f(s) = 0$.
\item[(iii)] $f$ is continuously differentiable with
$$f'(s) \ = \ \exp(a\,s-s^2/2)\,\bigl(b\,s - a\,b - \varphi(a)\bigr).$$
\end{enumerate}
By (iii) the function $f$ has a unique global minimum at $s_1 = \frac{\varphi(a)}{b} + a$.
By (ii) it must then hold that $f(s_1) < 0$. By continuity of $f$ and (i), there is $s_0 \in (0,s_1)$ such 
that $f(s_0) = 0$. This zero of $f$ is unique because $f$ has got only one extremum. \hfill\qed

By Proposition~\ref{pr:lognormal}, for each triple $(\alpha, q, t)$ with $0 < \alpha < 1$ and $0 < t < q$ there
is exactly one solution $(m,s) \in \mathbb{R} \times (0,\infty)$ of \eqref{eq:LognormalSolve} and therefore also
\eqref{eq:SysLognormal}. The component $s$ of this solution satisfies the inequality
\begin{equation}\label{eq:solution}
0  \ < \ s \ < \ \frac{\varphi\bigl(\Phi^{-1}(\alpha)\bigr)\,q}{\alpha\,t} + \Phi^{-1}(\alpha).
\end{equation}
The left-hand side of inequality \eqref{eq:solution} can be somewhat sharpened by making use of the fact that
by the definition of $\mathrm{ES}_\alpha$ and the first equation of \eqref{eq:LognormalSolve} we have
\begin{equation*}
t \ = \ ES_\alpha\bigl(\exp(m+s\,X)\bigr) \ < \ \mathrm{E}[\exp(m+s\,X)] \ = \ e^{m+s^2/2} 
\ = \ q \, e^{s^2/2- \Phi^{-1}(\alpha)\,s}.
\end{equation*}
Some algebra shows that in the case of $\bigl(\Phi^{-1}(\alpha)\bigr)^2 \ge 2\,\log(q/t)$ this implies
\begin{equation}
s \ > \ \Phi^{-1}(\alpha) + \sqrt{\bigl(\Phi^{-1}(\alpha)\bigr)^2 - 2\,\log(q/t)}.
\end{equation}
In the case of $\bigl(\Phi^{-1}(\alpha)\bigr)^2 < 2\,\log(q/t)$ there is no further constraint for $s$, i.e.\ 
we just have $s >0$.

Another sharper lower bound for $s$ (under a condition on the given quantile $q$ and ES $t$) can be concluded from  
the following theorem which formalises an observation made by \citet{Das&Stein} in the context of
securitisation.

\begin{theorem}\label{th:convex}
Let $Z$ be real random variable and $q > 0$ such that $\mathrm{P}[Z \le 0] = 0$, $\mathrm{P}[Z \le q] > 0$ and 
$z \mapsto \mathrm{P}[Z \le z]$ is convex for $z \in (0,q]$. Then it holds that
$$\mathrm{E}[Z\,|\,Z \le q] \ \ge \ \frac q 2.$$
\end{theorem}
\textbf{Proof.} Observe that 
\begin{equation}\label{eq:aux}
    \mathrm{E}[Z\,\mathbf{1}_{\{Z\le q\}}]\ =\ q\,\mathrm{P}[Z \le q] - \int_0^q \mathrm{P}[Z \le z]\,dz.
\end{equation}
For any $0< r < q$ and $r \le z \le q$ we have by the convexity of $z \mapsto \mathrm{P}[Z \le z]$ that
$$\mathrm{P}[Z \le z] \ \le \ \frac{q-z}{q-r}\,\mathrm{P}[Z \le r] + \frac{z-r}{q-r}\,\mathrm{P}[Z \le q].$$
By $r \to 0$ this implies for $0 < z \le q$ that
$$\mathrm{P}[Z \le z] \ \le \ \frac z q\,\mathrm{P}[Z \le q].$$
Making use of this inequality in \eqref{eq:aux} gives
\begin{align*}
    \mathrm{E}[Z\,\mathbf{1}_{\{Z\le q\}}] &\ \ge\ q\,\mathrm{P}[Z \le q] - \frac{q\,\mathrm{P}[Z \le q]}2\\
        & \ = \  \frac{q\,\mathrm{P}[Z \le q]}2.
\end{align*}
Dividing by $\mathrm{P}[Z \le q]$ now proves the assertion. \hfill\qed

Assume that in \eqref{eq:SysLognormal} we have $t < q/2$. 
Because lognormal densities are uni-modal Theorem~\ref{th:convex} then implies that the mode of the distribution is
smaller than $q$, i.e. 
$$e^{m-s^2} \ < \ q.$$
Making use of the first equation of \eqref{eq:LognormalSolve} this gives
\begin{equation}
q\,e^{-\Phi^{-1}(\alpha)\,s - s^2} \ < \ q \quad
\iff \quad s \  > \ -\Phi^{-1}(\alpha).
\end{equation}
For $\alpha \le 0.001$ hence $s$ will necessarily be greater than 3 if the condition $t < q/2$ on the
input data holds.

More generally, in terms of recovery rates and LGDs, Theorem~\ref{th:convex} suggests that for
'intuitive' asset value distributions with unimodal densities the expected recovery rate $RR$ is not
smaller than 50\% (or, equivalently, the LGD is not greater than 50\%) as long as the default threshold $D$
is not greater than the mode of the asset value density.


\section{A simple model of covered bonds losses}
\label{se:CovBonds}

There is not much literature on mathematical pricing or credit loss models for covered bonds. 
\begin{itemize}
\item A detailed pricing model based on time-continuous modelling of defaults and 
replacements of defaulted assets in the cover pool is presented by \citet{kenyon2009pricing}. 
\item A one-period model for recovery rates from a bank portfolio consisting of
one pool of encumbered assets and one pool of unencumbered assets is discussed by
\citet[][Section~II~D]{chan-lau&Oura2014bail}. \citeauthor{chan-lau&Oura2014bail} model the value 
of the whole bank portfolio by one random variable. This implies an assumption of homogeneity between the 
encumbered and the unencumbered pools such that both sub-portfolios can be modelled 
with a single asset value variable.
\end{itemize}
Driven by the problems with calibration of the model in the absence of proper calibration data (i.e.\ default
and loss observations for covered bonds), we follow here the simple 
approach of \citet{chan-lau&Oura2014bail} but introduce several refinements:
\begin{itemize}
\item We start in this section with a model with two asset value variables: 
one for the encumbered pool (e.g.\ the cover pool for
covered bonds) and one for the unencumbered pool, i.e.\ the part of the portfolio which complements the
cover pool.
\item We then assume that the bank's total debt consists of three components: secured (by the cover pool, overcollateralised) debt,
senior unsecured debt, and junior unsecured debt. 
Covered bonds usually are senior secured debt: In case of the cover pool being insufficient for the service of the 
covered bonds, the uncovered claim from the covered bonds debt is pari passu with the senior unsecured debt and higher-ranking
than the junior unsecured debt.
\item In addition, in Section~\ref{se:1lognormal} we refine the one-asset model of 
\citet{chan-lau&Oura2014bail}
by an adjustment of the encumbrance ratio in order to take into account the risk profile of the cover pool and
also study the three components debt structure described above in the two-assets case.
\end{itemize}

\subsection{One-period, two assets approach}
\label{se:2assets}

\begin{figure}[t!p]
\caption{Illustration of Equation~\eqref{eq:ev3}: Figure~10 of \citet{chan-lau&Oura2014bail}.}
\label{fig:10}
\begin{center}
\ifpdf
	\includegraphics[width=12cm]{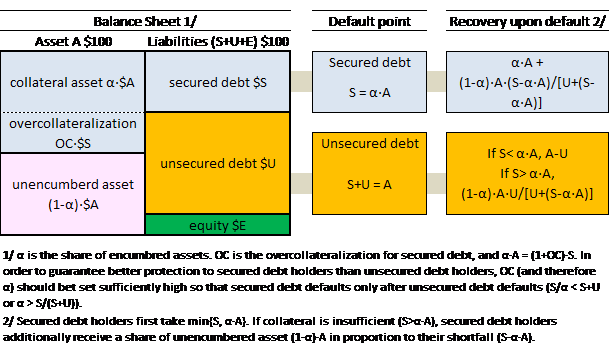}
\fi
\end{center}
\end{figure}

In this paper, we only study the case where both senior and junior (i.e.\ subordinated) debt are 
present in the portfolio of the covered bonds issuer. 
From the formulae of that case, the formulae for the
case without junior debt are readily obtained by setting $S=U$ and $U=0$ in the following equations.
In mathematical terms, the setting is as follows:
\begin{subequations}
\begin{itemize}
\item $C$ is the amount of senior secured (by the cover pool) debt in the issuer's balance sheet. 
$S$ is the amount of senior unsecured debt. 
$U$ is the amount of remaining debt, assumed to be subordinated unsecured debt.
\item $v$ is the level of over-collateralization of the covered bonds. $v=20\%$ means that the value of the 
collateral is 120\% of the face value of the covered bonds. 
For regulated covered bonds, in general there is a minimum requirement of over-collateralisation. 
Below, the parameter $v$ is used for model calibration.
\item $X$ is the future value of the encumbered cover pool assets (securing the covered bonds), 
$Y$ is the future value of the remaining assets, 
and $Z = X+Y$ is the future total value of the issuer's portfolio. As asset values, $X$ and $Y$ cannot
be negative.
We consider the values of $X$, $Y$ and $Z$ as
unpredictable and, therefore, treat them as random variables. 
\item $R_C$ is the recovery rate of the covered bonds, $R_S$ is the recovery rate of the senior unsecured 
debt, $R_U$ is the recovery rate of the junior unsecured debt. $R_C$, $R_S$ and $R_U$ are random variables.
\item $L_C = 1 - R_C$ is the loss rate of the covered bonds, $L_S = 1 -R_S$ is the loss rate of the 
senior unsecured debt, $L_U = 1 - R_U$ is the loss rate of the junior unsecured debt.
\end{itemize}
In the simplifying one-period context, the following events at the end of the observation 
period (described in mathematical terms) cause financial losses to the 
creditors of the junior debt, the senior unsecured debt or the covered bonds:
\begin{itemize}
\item The issuer defaults, but the total assets suffice to pay for all senior debt 
(covered bonds and unsecured senior debt): 
$$Z < C+S+U,\quad Z \ge C+S.$$ 
The loss rates experienced by the different classes of creditors then are as follows: 
\begin{equation}\label{eq:ev1}
L_U \ = \ 1 - \frac{Z-(C+S)}U, \qquad L_C \ = L_S \ = \ 0.
\end{equation}
\item The issuer defaults, the total assets do not suffice to pay for all senior debt (covered 
bonds and unsecured senior debt). The cover pool suffices to pay for the debt service of covered bonds.
We assume that the surplus from
the cover pool (i.e.\ the amount of cover pool assets not needed to pay for the covered bonds) is given 
to the creditors of the senior unsecured debt (which, in practice, might only happen after many years). This
scenario can be described as 
$$Z < C+S+U,\quad Z < C+S,\quad X \ge C.$$ 
The implied loss rate are as follows:
\begin{equation}\label{eq:ev2}
L_C \ = \ 0, \qquad L_S \ = \ 1 - \frac{Z-C}S, \qquad L_U \ = \ 1.
\end{equation}
\item The issuer defaults, the total assets do not suffice to pay for all senior debt (covered 
bonds and unsecured senior debt). The cover pool does not suffice to pay for the debt service of 
the covered bonds:
\begin{equation}\label{eq:LossEvent}
Z < C+S+U,\quad Z < C+S,\quad X < C.
\end{equation}
In this scenario the creditors experience these loss rates 
\citep[][see Figure~\ref{fig:10}]{chan-lau&Oura2014bail}:
\begin{equation}\label{eq:ev3}
L_C \ = \ \frac{(C-X)\,(S+C-Z)}{(S+C-X)\,C}, \quad
L_S \ = \ \frac{S+C-Z}{S+C-X}, \quad L_U \ = \ 1.
\end{equation}
\end{itemize}
\end{subequations}
At first glance, the case where in total the value of the issuer's assets is sufficient to pay all debt, 
but due to the asset encumbrance by the covered bonds there is not enough liquidity to pay for the 
unsecured debt ($Z \ge C+S+U$, $X \ge C$, $Y < S+U$) also might be considered a loss event. 
For the purpose of this paper, however, we do not treat this event as a loss event 
because in the longer term the creditors of the unsecured debt should be fully paid 
from the surplus of the cover pool.

Similarly, the event $Z \ge C+S+U$, $X < C$ is not a loss event either because the covered bonds holders' 
first claim is against the bond issuer which is solvent although the cover pool 
(despite over-collateralisation) would be insufficient to pay for the covered bonds.

\begin{remark}\label{rm:consequence}
An important consequence of the loss assumptions \eqref{eq:ev1}, \eqref{eq:ev2}, and \eqref{eq:ev3} is
that the bond issuer suffers a loss (i.e.\ the issuer defaults) if and only if 
$$Z \ = \ X + Y \ < \ C +S + U.$$
Hence for the PD $\mathrm{P}[Z < C+S+U]$ of the bond issuer it holds that
\begin{equation}\label{eq:PD.restrict}
\mathrm{P}[Z < C+S+U] \ \le \ \mathrm{P}[X < C+S+U].
\end{equation} 
If the credit quality of the cover pool assets is much better than the credit quality
of the bond issuer's remaining assets, \eqref{eq:PD.restrict} can be a major restriction for the two-assets
model because it implies that the issuer's PD cannot become arbitrarily bad 
(see Section~\ref{se:CaseLognormal} for an illustrative example). To be more precise:
Note that by \eqref{eq:LossEvent}, the 
standalone (i.e.\ without support by the bond issuer) probability of loss of the cover pool is
$\mathrm{P}[X < C]$. Hence, if the credit quality of the cover pool is good (i.e.\ $\mathrm{P}[X < C]$
is small) and $S+U$ is small compared to $C$ (i.e.\ most of the firm's portfolio is pledged as 
collateral for the covered bonds) then \eqref{eq:PD.restrict} implies that the bond issuer's PD is also 
small even if the other assets are very risky.

In any case, it is important to keep in mind that the probabilities in \eqref{eq:PD.restrict} may refer
to different time horizons, depending on the purpose the model is deployed for:
\begin{itemize}
\item If the model is used for calculating (say) one-year PD and LGD for the bond issuer, $\mathrm{P}[Z < C+S+U]$
refers to the issuer's one-year PD while $\mathrm{P}[X < C+S+U]$ should be related to the maturity of the
covered bonds to which the cover pool assets serve as collateral. For only after the cover pool assets are no
longer pledged it will become clear to which extent the issuer's unsecured creditors can take recourse to
the cover pool assets to satisfy their claims.
\item If the model is used for calculating (say) one-year expected loss for the covered bonds, both
$\mathrm{P}[Z < C+S+U]$ and $\mathrm{P}[X < C+S+U]$ must refer to a one-year time horizon. For in this
case a covered bonds loss event can only occur if the bond issuer defaults. Default of the covered bonds
then is triggered by the value of the cover pool after one year. The covered bonds recovery value after one year is
determined by the one-year values of the cover pool assets and the remaining assets.
\end{itemize}
\end{remark}

\subsection{Covered bonds expected loss in the two-assets lognormal model}
\label{se:2assetsEL}
Ever since the seminal paper by \citet{Merton1974} was published, modelling of a company's asset value distribution
by a lognormal distribution has been very popular. To do so is convenient if not always realistic. Here, we follow this
approach mainly because it allows fitting a model to sparse given data.

Assume that $X$ and $Y$ from Section~\ref{se:2assets} are both lognormal variables, linked by a normal copula.
Hence 
\begin{equation}\label{eq:lognormal}
X \ = \ \exp(\mu + \sigma\,\xi), \qquad Y \ = \ \exp(\nu + \tau\,\eta),
\end{equation}
with $\mu, \nu \in \mathbb{R}$, $\sigma, \tau > 0$ and $(\xi, \eta) \sim \mathcal{N}\left(
\begin{pmatrix}
0 \\ 0
\end{pmatrix}, \begin{pmatrix}
1 & \varrho \\
\varrho & 1
\end{pmatrix}\right)$ for some $\varrho \in [0,1]$.
Since the sum of two lognormal random variables in general is not lognormally 
distributed, most of the probabilities and expectations of the variables listed in Section~\ref{se:2assets} 
can only be numerically evaluated. To facilitate the numerical
calculations, in the following, we provide formulae for the evaluation of the loss probabilities and 
expected losses that do not require more numerical effort than one-dimensional integration. 
We need to distinguish the two cases $\varrho < 1$ and $\varrho = 1$ (when $X$ and $Y$ are comonotonic).

\paragraph{Case $\varrho < 1$.} 
We make use of the fact that by the assumption of a normal copula linking
$\xi$ and $\eta$ in \eqref{eq:lognormal},
the distribution of $\eta$ conditional on $\xi=x$ is normal with mean $\varrho\,x$ and standard 
deviation $\sqrt{1-\varrho^2}$. By applying the disintegration theorem, we then obtain the following 
equations for the probabilities of the loss events listed in Section~\ref{se:2assets}:
\begin{subequations}
\begin{align}
\mathrm{P}[Z < C+S+U, Z \ge C+S] & = \mathrm{E}\bigl[\mathrm{P}[Y < C+S+U-X, Y \ge C+S-X \,|\,X]\bigr]\notag\\
& = \int\limits_{-\infty}^{\frac{\log(C+S+U)-\mu}{\sigma}}
\varphi(x)\,\Phi\left(\tfrac{\log(C+S+U-e^{\mu+\sigma\,x})-(\nu+\tau\,\varrho\,x)}{\tau\,\sqrt{1-\varrho^2}}\right) dx  \notag\\
& \qquad  -\int\limits_{-\infty}^{\frac{\log(C+S)-\mu}{\sigma}}
\varphi(x)\,\Phi\left(\tfrac{\log(C+S-e^{\mu+\sigma\,x})-(\nu+\tau\,\varrho\,x)}{\tau\,\sqrt{1-\varrho^2}}\right) dx,\\
\mathrm{P}[Z < C+S+U, Z < C+S, X \ge C] & = \int\limits_{\frac{\log(C)-\mu}{\sigma}}^{\frac{\log(C+S)-\mu}{\sigma}}
\varphi(x)\,\Phi\left(\tfrac{\log(C+S-e^{\mu+\sigma\,x})-(\nu+\tau\,\varrho\,x)}{\tau\,\sqrt{1-\varrho^2}}\right)\,dx,\\
\mathrm{P}[Z < C+S+U, Z < C+S, X < C] & = \int\limits^{\frac{\log(C)-\mu}{\sigma}}_{-\infty}
\varphi(x)\,\Phi\left(\tfrac{\log(C+S-e^{\mu+\sigma\,x})-(\nu+\tau\,\varrho\,x)}{\tau\,\sqrt{1-\varrho^2}}\right)\,dx.
\end{align}
\end{subequations}
We obtain formulae for the expected loss rates by integrating the loss variables from equations
\eqref{eq:ev1}, \eqref{eq:ev2} and \eqref{eq:ev3} over the respective loss events and then adding up
the results. In the derivation of \eqref{eq:loss1} and \eqref{eq:loss2}, 
we make use of the second equation in
\eqref{eq:SysLognormal}.
\begin{subequations}
\begin{multline}\label{eq:loss1}
C\,\mathrm{E}[L_C] \ =\ \int\limits^{\frac{\log(C)-\mu}{\sigma}}_{-\infty} \bigl(C-e^{\mu+\sigma\,x}\bigr)\,
\varphi(x)\,\Phi\left(\tfrac{\log(C+S-e^{\mu+\sigma\,x})-(\nu+\tau\,\varrho\,x)}{\tau\,\sqrt{1-\varrho^2}}\right) dx\\
 - e^{\nu+\tau^2\,(1-\varrho^2)/2} \int\limits^{\frac{\log(C)-\mu}{\sigma}}_{-\infty}
\frac{(C-e^{\mu+\sigma\,x})\,e^{\tau\,\varrho\,x}}{C+S-e^{\mu+\sigma\,x}}\,
\varphi(x)\,\Phi\left(\tfrac{\log(C+S-e^{\mu+\sigma\,x})-(\nu+\tau\,\varrho\,x)}{\tau\,\sqrt{1-\varrho^2}}
    - \tau\,\sqrt{1-\varrho^2}\right)\,dx,
\end{multline}
\begin{multline}\label{eq:loss2}
S\,\mathrm{E}[L_S] \ =\ (C+S)\,\mathrm{P}[Z < C+S, X \ge C]  -\int\limits_{\frac{\log(C)-\mu}{\sigma}}^{\frac{\log(C+S)-\mu}{\sigma}}
e^{\mu+\sigma\,x}\,
\varphi(x)\,\Phi\left(\tfrac{\log(C+S-e^{\mu+\sigma\,x})-(\nu+\tau\,\varrho\,x)}{\tau\,\sqrt{1-\varrho^2}}\right) dx\\
-\int\limits_{-\infty}^{\frac{\log(S)-\nu}{\tau}}
e^{\nu+\tau\,x}\,
\varphi(x)\left\{\Phi\left(\tfrac{\log(C+S-e^{\nu+\tau\,x})-(\mu+\sigma\,\varrho\,x)}{\sigma\,\sqrt{1-\varrho^2}}\right)- 
\Phi\left(\tfrac{\log(C)-(\mu+\sigma\,\varrho\,x)}{\sigma\,\sqrt{1-\varrho^2}}\right)\right\} dx\\
 - S\,e^{\nu+\tau^2\,(1-\varrho^2)/2} \int\limits^{\frac{\log(C)-\mu}{\sigma}}_{-\infty}
\frac{e^{\tau\,\varrho\,x}}{C+S-e^{\mu+\sigma\,x}}\,
\varphi(x)\,\Phi\left(\tfrac{\log(C+S-e^{\mu+\sigma\,x})-(\nu+\tau\,\varrho\,x)}{\tau\,\sqrt{1-\varrho^2}}
    - \tau\,\sqrt{1-\varrho^2}\right) dx\\
    + S\,\mathrm{P}[Z < C+S,\, X < C],
\end{multline}
\begin{multline}\label{eq:loss3}
U\,\mathrm{E}[L_U] \ =\ U\,\mathrm{P}[Z < C+S] + 
	(C+S+U)\,\mathrm{P}[C+S \le Z < C+S+ U]\\ + 
\int\limits_{-\infty}^{\frac{\log(C+S)-\mu}{\sigma}}
e^{\mu+\sigma\,x}\,
\varphi(x)\,\Phi\left(\tfrac{\log(C+S-e^{\mu+\sigma\,x})-(\nu+\tau\,\varrho\,x)}{\tau\,\sqrt{1-\varrho^2}}\right) dx\\
- \int\limits_{-\infty}^{\frac{\log(C+S+U)-\mu}{\sigma}}
e^{\mu+\sigma\,x}\,
\varphi(x)\,\Phi\left(\tfrac{\log(C+S+U-e^{\mu+\sigma\,x})-(\nu+\tau\,\varrho\,x)}{\tau\,\sqrt{1-\varrho^2}}\right) dx\\
+ \int\limits_{-\infty}^{\frac{\log(C+S)-\nu}{\tau}}
e^{\nu+\tau\,x}\,
\varphi(x)\,\Phi\left(\tfrac{\log(C+S-e^{\nu+\tau\,x})-(\mu+\sigma\,\varrho\,x)}{\sigma\,\sqrt{1-\varrho^2}}\right) dx\\
-\int\limits_{-\infty}^{\frac{\log(C+S+U)-\nu}{\tau}}
e^{\nu+\tau\,x}\,
\varphi(x)\,\Phi\left(\tfrac{\log(C+S+U-e^{\nu+\tau\,x})-(\mu+\sigma\,\varrho\,x)}{\sigma\,\sqrt{1-\varrho^2}}\right) dx.
\end{multline}
\end{subequations}
\paragraph{Case $\varrho = 1$.} In this case, the asset value variables $X$ and $Y$ are comonotonic, i.e.\
if one of them is known the value of the other is also known. This is the strongest possible type of 
dependence between two random variables. It defines a worst case scenario because high losses with
the cover pool of the covered bonds occur exactly at the same time when also the losses with the issuer's
remaining assets are high. Denote by $x(a)$ for $a>0$ the unique solution of the
equation
\begin{equation}\label{eq:uniquesol}
 a \ = \ e^{\mu+\sigma\,x} + e^{\nu+\tau\,x}.
\end{equation}
The probabilities of the loss events from Section~\ref{se:2assets} then can be calculated as follows
\begin{subequations}
\begin{align}
\mathrm{P}[Z < C+S+U, Z \ge C+S]  & =
\Phi\bigl(x(C+S+U)\bigr) - \Phi\bigl(x(C+S)\bigr),\label{eq:p1}\\
\mathrm{P}[Z < C+S+U, Z < C+S, X \ge C]  & =
	\max\big(\Phi\bigl(x(C+S)\bigr) - \Phi\bigl(\tfrac{\log(C)-\mu}{\sigma}\bigr), 0\bigr),\label{eq:p2}\\
\mathrm{P}[Z < C+S+U, Z < C+S, X < C] & =
		\Phi\bigl(\min\bigl(x(C+S),\tfrac{\log(C)-\mu}{\sigma}\bigr)\bigr).\label{eq:p3}
\end{align}
\end{subequations}
Although in this case the model is basically one-dimensional, due to the non-linear structure of the loss variables
in \eqref{eq:ev3}, numerical integration cannot be avoided when evaluating the following equations 
\eqref{eq:com1} and \eqref{eq:com2} for the expected losses of the covered bonds and senior unsecured debt
holders. In contrast, expected loss for the junior debt holders does not require much computational effort, as
shown in \eqref{eq:com3}.
\begin{subequations}
\begin{equation}\label{eq:com1}
C\,\mathrm{E}[L_C]  = 
	 \int\limits_{-\infty}^{\min\left(\frac{\log(C)-\mu}{\sigma},\, x(C+S)\right)}
	\varphi(x)\,\frac{\bigl(C-e^{\mu+\sigma\,x}\bigr) \bigl(C+S-e^{\mu+\sigma\,x}-e^{\nu+\tau\,x}\bigr)}
	{C+S-e^{\mu+\sigma\,x}}\,dx,
\end{equation}
\begin{multline}\label{eq:com2}
S\,\mathrm{E}[L_S]  = S\,\mathrm{P}[Z < C+S, X < C] -
	S \int\limits_{-\infty}^{\min\left(\frac{\log(C)-\mu}{\sigma},\, x(C+S)\right)}
	\varphi(x)\,\frac{e^{\nu+\tau\,x}}
	{C+S-e^{\mu+\sigma\,x}}\,dx \\
	+ (C+S)\,\mathrm{P}[Z < C+S, X \ge C] -
	e^{\mu+\sigma^2/2}\,\max\big(\Phi\bigl(x(C+S)-\sigma\bigr) - \Phi\bigl(\tfrac{\log(C)-\mu}{\sigma}-\sigma\bigr), 0\bigr)\\
	- e^{\nu+\tau^2/2}\,\max\big(\Phi\bigl(x(C+S)-\tau\bigr) - \Phi\bigl(\tfrac{\log(C)-\mu}{\sigma}-\tau\bigr), 0\bigr),
\end{multline}
\begin{multline}\label{eq:com3}
U\,\mathrm{E}[L_U]  = 
	U\,\mathrm{P}[Z < C+S] + (C+S+U)\,\mathrm{P}[Z < C+S+U, Z \ge C+S] \\
	- e^{\mu+\sigma^2/2}\,\left(\Phi\bigl(x(C+S+U)-\sigma\bigr) - \Phi\bigl(x(C+S)-\sigma\bigr)\right) \\
	- e^{\nu+\tau^2/2}\,\left(\Phi\bigl(x(C+S+U)-\tau\bigr) - \Phi\bigl(x(C+S)-\tau\bigr)\right).
\end{multline}
\end{subequations}

\section{Calibration of the two-assets model}
\label{se:2lognormal}

The credit risk of a financial institution typically is stated in terms of the institution's PD and LGD. As
described in Section~\ref{se:VaRES}, then often  a distribution of the institution's assets value 
can be fitted by a variation of the 
method of moments such that the given PD and LGD are matched (see \eqref{eq:PD} and \eqref{eq:RR}). 
However, in order to implement a 
two-assets model like in Sections \ref{se:2assets} and \ref{se:2assetsEL}, we need the joint distribution of
the pair $(X,Y)$ of the values of the cover pool and of the remaining assets respectively. 

\subsection{The lognormal case}
\label{se:CaseLognormal}

In the case of the two-assets
lognormal model as specified by \eqref{eq:lognormal}, we have to determine the values of the five 
parameters $\mu, \sigma$, $\nu, \tau$ and $\varrho$.
It would be nice if there were a way to split the PD and LGD (or equivalently the PD and expected loss) 
associated with
a financial institution that issues covered bonds into separate PD and LGD estimates for the cover pool and 
the pool of the remaining assets respectively. 
Then we could use the approach from Section~\ref{se:ESlognormal} to
fit lognormal marginal distributions of the cover pool assets and other assets respectively (i.e.\ the 
parameters $\mu, \sigma$, $\nu, \tau$) and either try to somehow estimate the correlation $\varrho$ or just
choose an appropriate value for $\varrho$. As there is an obvious worst case (the case $\varrho = 1$ with
comonotonic marginal distributions), choosing $\varrho$ might be a better approach than estimating it 
because this way the model easily can be made to err on the conservative side.

Without further constraints, it is likely that there are many ways of splitting the issuer's PD and LGD into
PD and LGD parameters of the sub-portfolios, and some of these ways might even be reasonable from an
economic perspective. But some constraints actually make sense. Given that regulatory approval for a covered 
bonds issuance is required if relief of capital requirements for the bond investors is intended, the cover 
pool is likely to be subject to a number of rules regarding its composition and riskiness. Therefore, 
it seems safe to 
assume that probability of loss and expected loss estimates $p_{\mathrm{cover}}$ and 
$EL_{\mathrm{cover}} = p_{\mathrm{cover}}\,LGD_{\mathrm{cover}}$ for the cover pool are available.

A number of different approaches to the calibration of the cover pool asset value parameters to given
$p_{\mathrm{cover}}$ and 
$EL_{\mathrm{cover}}$ are conceivable. Here we suggest calibrating to the probability of the event that
the cover pool asset value falls below the threshold 'face value of the covered bonds times
(100\% plus level of over-collateralisation)' and the expected value of the shortfall below
the threshold.
Denote by $C > 0$ the face value of the covered bonds and 
by $v > 0$ the level of 
over-collateralisation of the covered bonds, as in Section~\ref{se:2assets}. Hence we have 
$$\text{Face value of the covered bonds} \times
\text{(100\% plus level of over-collateralisation)} \ = \ C\,(1+v).$$

For determining the parameters  of the lognormal 
representation \eqref{eq:lognormal} of the cover pool asset value we make use of the observation in 
Remark~\ref{rm:PDvQuant} about the relationship between PD, threshold and recovery rate on the one hand and
PD, quantile and ES on the other hand. Once we have expressed quantile and ES of the asset value in
terms of the threshold $C\,(1+v)$ and $EL_{\mathrm{cover}} = p_{\mathrm{cover}}\,(1-RR_{\mathrm{cover}})$, 
we can use results from Section~\ref{se:ESlognormal} to infer the values of $\mu$ and $\sigma$.

Let $X$ denote the asset value of the cover pool. As suggested in Remark~\ref{rm:PDvQuant}, we then
have $C\,(1+v) = q_{p_{\mathrm{cover}}}(X)$ and
$$
\frac{ES_{p_\mathrm{cover}}(X)}{C\,(1+v)}  \ = \ RR_{\mathrm{cover}}\quad
\iff \quad ES_{p_\mathrm{cover}}(X) \ = \ 
	C\,(1+v)\,\frac{p_{\mathrm{cover}}-EL_{\mathrm{cover}}}{p_{\mathrm{cover}}}.
$$
\eqref{eq:SysLognormal} and \eqref{eq:LognormalSolve} now imply the following 
equation system for $\mu$ and $\sigma$:
\begin{equation}\label{eq:coverpool}
\begin{split}
\mu & \ = \ \log((1+v)\, C) - \sigma\,\Phi^{-1}(p_{\mathrm{cover}}),\\
0 & \ = \ \Phi\bigl(\Phi^{-1}(p_{\mathrm{cover}})-\sigma\bigr) -
		 \bigl(p_{\mathrm{cover}}-EL_{\mathrm{cover}}\bigr)\,
		 \exp\bigl(\sigma\,\Phi^{-1}(p_{\mathrm{cover}})-\sigma^2/2\big).
\end{split}
\end{equation}
The factor $1+v$ in the first row of 
\eqref{eq:coverpool} reflects the fact that the risk characteristics are provided for the whole cover pool,
not only for that part of it which has exactly the value of the covered bonds.
Recall that according to Proposition~\ref{pr:lognormal}, \eqref{eq:coverpool} has a unique solution
$(\mu, \sigma) \in \mathbb{R}\times (0, \infty)$. Parameters $\mu$ and $\sigma$ of the cover pool asset value
hence can be assumed to be known. This is not the case for the parameters $\nu$ and $\tau$ of the distribution 
of the value of
the remaining assets in the bond issuer's portfolio because typically no detailed information is 
publicly available
about the structure and riskiness of a bank's portfolio unless the assets are pledged as collateral for
borrowing. 

Nonetheless, we can try and determine implied values for $\nu$ and $\tau$ as follows. Denote by 
$p_{\mathrm{issuer}}$ the issuer's PD, by $LGD_{\mathrm{issuer}}$ the issuer's LGD\footnote{%
For the sake of simplicity, we assume that we are given an LGD for the issuer's entire debt. In practice, it
might be more likely to have an LGD for the issuer's senior unsecured debt. This case could be dealt with by replacing
the right-hand side of \eqref{eq:2optim2} with the right-hand side of \eqref{eq:loss2} in case of $\varrho < 1$ 
and the right-hand side
of \eqref{eq:2com2} with the right-hand side of \eqref{eq:com2} in case of $\varrho = 1$.} and by 
$D_{\mathrm{issuer}}$ the issuer's total amount of outstanding debt (including the covered bonds issued). 
In the setting of Section~\ref{se:2assets}, it would hold that $D_{\mathrm{issuer}} = C + S + U$.
As in Section~\ref{se:2assetsEL} we then represent $p_{\mathrm{issuer}}$ and $LGD_{\mathrm{issuer}}$ as 
functions of the parameters $\mu, \sigma$, $\nu, \tau$ and $\varrho$. 

In case of $\varrho < 1$, with $X$ and $Y$ as in 
\eqref{eq:lognormal}, we obtain as the first equation
\begin{subequations}
\begin{align}
	p_{\mathrm{issuer}} & = \mathrm{P}[X+Y < D_{\mathrm{issuer}}]\notag \\
	& = \int\limits_{-\infty}^{\frac{\log(D_{\mathrm{issuer}})-\mu}{\sigma}}
\varphi(x)\,\Phi\left(\tfrac{\log(D_{\mathrm{issuer}}-e^{\mu+\sigma\,x})-(\nu+\tau\,\varrho\,x)}
	{\tau\,\sqrt{1-\varrho^2}}\right)\,dx.\label{eq:2optim1}
\end{align}
For the second equation, we obtain
\begin{multline}
	D_{\mathrm{issuer}}\,p_{\mathrm{issuer}}\,(1-LGD_{\mathrm{issuer}})\  =\
	\mathrm{E}\bigl[(X+Y)\,\mathbf{1}_{\{X+Y < D_{\mathrm{issuer}}\}}\bigr] \\
	\ =\ \int\limits_{-\infty}^{\frac{\log(D_{\mathrm{issuer}})-\mu}{\sigma}}
e^{\mu+\sigma\,x}\,
\varphi(x)\,\Phi\left(\tfrac{\log(D_{\mathrm{issuer}}-e^{\mu+\sigma\,x})-(\nu+\tau\,\varrho\,x)}{\tau\,\sqrt{1-\varrho^2}}\right) dx\\
  + e^{\nu+\tau^2\,(1-\varrho^2)/2} \int\limits_{-\infty}^{\frac{\log(D_{\mathrm{issuer}})-\mu}{\sigma}}
e^{\tau\,\varrho\,x}\,
\varphi(x)\,\Phi\left(\tfrac{\log(D_{\mathrm{issuer}}-e^{\mu+\sigma\,x})-(\nu+\tau\,\varrho\,x)}{\tau\,\sqrt{1-\varrho^2}}
    - \tau\,\sqrt{1-\varrho^2}\right) dx.\label{eq:2optim2}
\end{multline}
\end{subequations}
In case of $\varrho = 1$, the first equation is
\begin{subequations}
\begin{equation}\label{eq:2com1}
p_{\mathrm{issuer}} \  = \ \Phi\bigl(x(D_{\mathrm{issuer}})\bigr),
\end{equation}
where for $a > 0$ the function $x(a)$ is defined as the solution of \eqref{eq:uniquesol}. The second equation is given
by
\begin{equation}\label{eq:2com2}
D_{\mathrm{issuer}}\,p_{\mathrm{issuer}}\,(1-LGD_{\mathrm{issuer}})\  =\
e^{\mu + \sigma^2/2}\,\Phi\bigl(x(D_{\mathrm{issuer}}) - \sigma\bigr) +
e^{\nu + \tau^2/2}\,\Phi\bigl(x(D_{\mathrm{issuer}}) - \tau\bigr).
\end{equation}
\end{subequations}
Assuming that parameters $\mu$, $\sigma$ and $\varrho$ are known, we have to solve the equation system
\eqref{eq:2optim1} and \eqref{eq:2optim2} (or \eqref{eq:2com1} and \eqref{eq:2com2} respectively) 
for $\nu$ and $\tau$ in order to fully parameterise the lognormal
two-assets model from Section~\ref{se:2assetsEL}. Unfortunately, there is no obvious way to transform the equation
system \eqref{eq:2optim1} and \eqref{eq:2optim2} such that variables $\nu$ and $\tau$ are separated. This is
different for equation system \eqref{eq:2com1} and \eqref{eq:2com2} in the comonotonic case $\varrho = 1$. We can
therefore state the following result on the solutions of the system.
The proposition shows that there are combinations
of the input data $p_{\mathrm{cover}}$, $EL_{\mathrm{cover}}$, $C$, $v$,
$p_{\mathrm{issuer}}$, $LGD_{\mathrm{issuer}}$, and $D_{\mathrm{issuer}}$ such that the equation system
\eqref{eq:2com1} and \eqref{eq:2com2} for $\nu$ and $\tau$ has no solution at all. 
\begin{proposition}\label{pr:syssolution}
Assume that $D_{\mathrm{issuer}}> 0$, $0< p_{\mathrm{issuer}} < 1$, $0< LGD_{\mathrm{issuer}} < 1$, $\mu \in \mathbb{R}$ and
$\sigma > 0$ are known and kept fix. Then there exists a solution $(\nu, \tau) \in \mathbb{R} \times (0,\infty)$ 
of the equation system \eqref{eq:2com1} and \eqref{eq:2com2} if and
only if it holds that
\begin{subequations}
\begin{gather}
0 \ < \ p_{\mathrm{issuer}}\ < \ \Phi\left(\frac{\log(D_{\mathrm{issuer}})-\mu}{\sigma}\right) \quad\text{and}
\label{eq:p.restrict}\\
\frac{p_{\mathrm{issuer}}\,e^{\mu+\sigma\,\Phi^{-1}(p_{\mathrm{issuer}})}-
e^{\mu+\sigma^2/2}\,\Phi\bigl(\Phi^{-1}(p_{\mathrm{issuer}})-\sigma\bigr)}{p_{\mathrm{issuer}}\,D_{\mathrm{issuer}}}
 <  LGD_{\mathrm{issuer}}  <  1 - \frac{e^{\mu+\sigma^2/2}\,\Phi\bigl(\Phi^{-1}(p_{\mathrm{issuer}})-\sigma\bigr)}{p_{\mathrm{issuer}}\,D_{\mathrm{issuer}}}.\label{eq:LGD.restrict}
\end{gather}
\end{subequations}
If there is a solution $(\nu, \tau) \in \mathbb{R} \times (0,\infty)$  of \eqref{eq:2com1} and \eqref{eq:2com2} it 
is unique.
\end{proposition}
\textbf{Proof.} For the sake of a more concise notation, in this proof we drop the index 'issuer'. 
Observe that for any finite $(\nu, \tau)$ the solution $x(D)$ of the equation 
\begin{equation}\label{eq:constraint}
D\ =\ e^{\mu+\sigma\,x(D)} + e^{\nu+\tau\,x(D)}
\end{equation} 
must satisfy
$$x(D) \ < \ \frac{\log(D)-\mu}{\sigma}.$$
Comonotonicity ($\varrho = 1$) implies that the random variables $X$ and $Y$ from \eqref{eq:lognormal} can be represented
as $X = e^{\mu+\sigma\,\xi}$, $Y=e^{\nu+\tau\,\xi}$ for some standard normal variable $\xi$. Hence it follows that
$$p = \mathrm{P}[e^{\mu+\sigma\,\xi}+e^{\nu+\tau\,\xi} < D] = \mathrm{P}[\xi < x(D)] < 
\mathrm{P}\left[\xi < \frac{\log(D)-\mu}{\sigma}\right] =
\Phi\left(\frac{\log(D)-\mu}{\sigma}\right).$$
This shows that \eqref{eq:p.restrict} is necessary for the existence of a solution. 

Assume now that $p$ satisfies \eqref{eq:p.restrict} and
that $(\nu, \tau)$ is a solution of \eqref{eq:2com1} and \eqref{eq:2com2}. Then \eqref{eq:constraint} can be
solved for $\nu$:
$$\nu \  = \ \log\bigl(D - e^{\mu+\sigma\,x(D)}\bigr) - \tau\,x(D).$$ 
Substitute this for $\nu$ in the term $e^{\nu + \tau^2/2}\,\Phi\bigl(x(D) - \tau\bigr)$ of \eqref{eq:2com2} 
to obtain the term
$$g(\tau)\ \stackrel{\mathrm{def}}{=} \ \bigl(D - e^{\mu+\sigma\,x(D)}\bigr)\,
e^{\tau^2/2-\tau\,x(D)}\,\Phi\bigl(x(D) - \tau\bigr).$$
The derivative of $g$ with respect to $\tau$ is 
$$g'(\tau) \ =\ \bigl(D - e^{\mu+\sigma\,x(D)}\bigr)\,e^{\tau^2/2-\tau\,x(D)}\,
\bigl\{\Phi(x(D)-\tau)\,(\tau-x(D))-\varphi(\tau-x(D))\bigr\}.$$
By \eqref{eq:ineq}, if follows that $g'(\tau) < 0$ for all $\tau > 0$ such that $g$ is a strictly descreasing 
function. 
Hence $g$ is an 'onto' mapping from $(0, \infty)$ to 
$\bigl(\lim_{\tau\to\infty} g(\tau), \lim_{\tau\to 0} g(\tau)\bigr)$. It is easy to see (using again
\eqref{eq:ineq}) that
\begin{equation*}
\begin{split}
    \lim_{\tau\to 0} g(\tau) & = \bigl(D - e^{\mu+\sigma\,x(D)}\bigr)\,p,\\
    \lim_{\tau\to\infty} g(\tau) & = 0.
\end{split}
\end{equation*}
By rearranging \eqref{eq:2com2} such that $LGD$ is the only term on one side of the equation, it follows that
\eqref{eq:LGD.restrict} is necessary for the existence of a solution of \eqref{eq:2com1} and \eqref{eq:2com2}.
Sufficiency follows from the fact that $g$ is 'onto'. Uniqueness of the solution is a consequence of the fact
that $g$ is strictly decreasing. \hfill $\Box$

\begin{figure}[t!p]
\caption{Illustration of Proposition~\ref{pr:syssolution}. Ranges of bond issuer's PD and LGD for two
set of parameters for the cover pool. The smaller range (doubly shaded) is related to cover pool PD and LGD of
0.05\% and 30\% respectively. The larger range (singly shaded) is related to cover pool PD and LGD of
0.5\% and 50\% respectively. See Example~\ref{ex:ranges} for further explanation.}
\label{fig:1}
\begin{center}
\ifpdf
	\includegraphics[width=15cm]{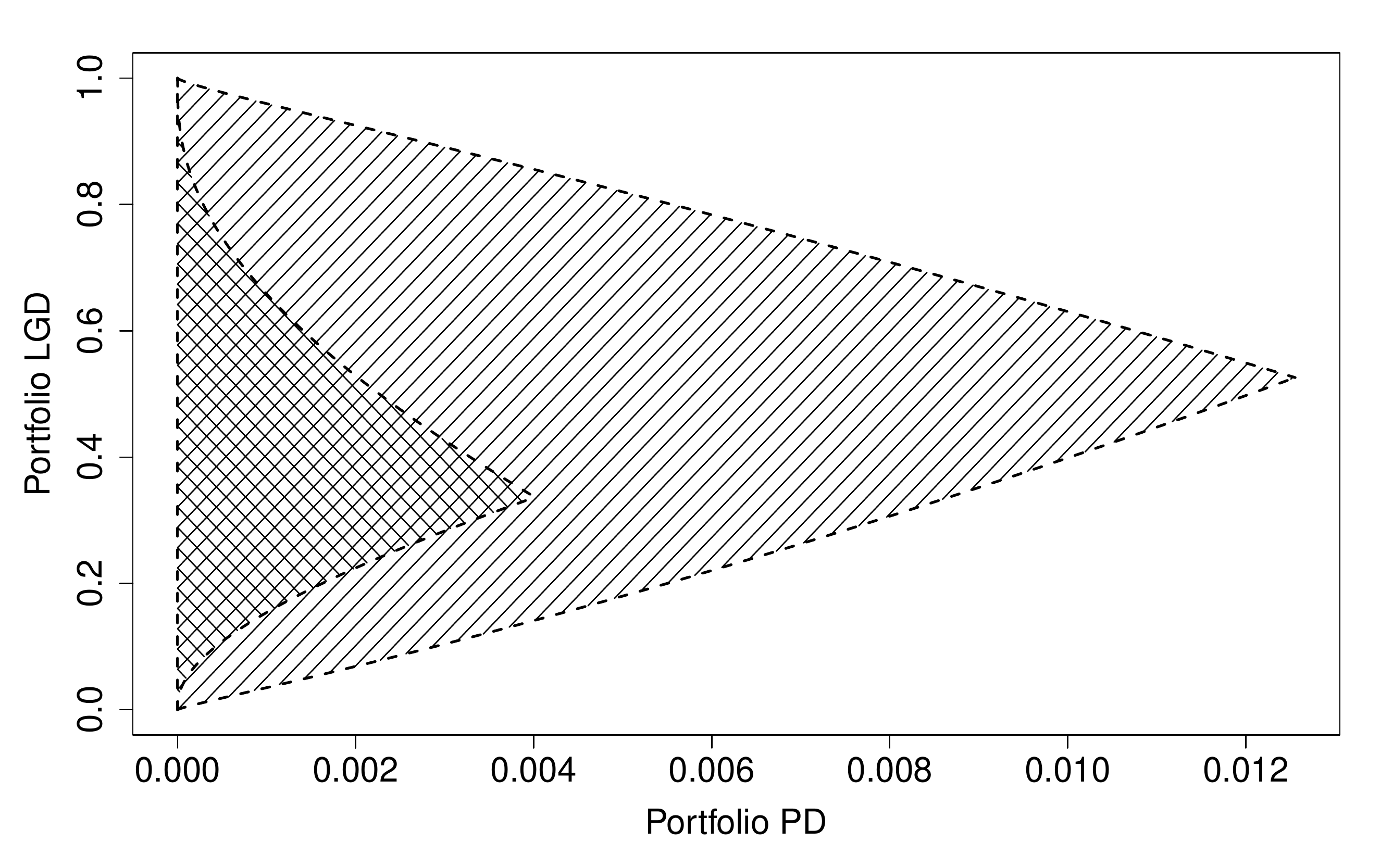}
\fi
\end{center}
\end{figure}

\begin{example}\label{ex:ranges}
To illustrate the consequences of Proposition~\ref{pr:syssolution}, in Figure~\ref{fig:1} we plot the
possible PD-LGD ranges of the bond issuer for two sets of parameters for the cover pool
(see Section~\ref{se:2assets} for information on the meaning of the parameters):
\begin{itemize}
\item $C=0.3$, $S=0.6$, $U=0.1$, $v=0.2$,
\item $(p_{\mathrm{cover}}, LGD_{\mathrm{cover}})  \in \{(0.0005, 0.3), (0.005, 0.5)\}$, 
\item $\varrho = 1$ (comonotonic case).
\end{itemize}
Reading example for Figure~\ref{fig:1}: $PD = 2\%$ for the bond issuer cannot be realised with either of the two parameter sets for
the cover pool. $PD = 0.8\%$, $LGD = 40\%$ for the bond issuer can be realised with 
$(p_{\mathrm{cover}}, LGD_{\mathrm{cover}})  = (0.005, 0.5)$ but not with 
$(p_{\mathrm{cover}}, LGD_{\mathrm{cover}})  = (0.0005, 0.3)$.
\end{example}

Intuitively, the restricted PD-LGD range for the bond issuer from
Proposition~\ref{pr:syssolution} and Example~\ref{ex:ranges} is caused by the 
following phenomenon: The input parameters $p_{\mathrm{cover}}$, $EL_{\mathrm{cover}}$,
$C$, and $v$ for the distribution of the cover pool asset value $X$ may force $X$ to have a 
large variance while at the same time the input parameters $p_{\mathrm{issuer}}$, $LGD_{\mathrm{issuer}}$, 
and $D_{\mathrm{issuer}}$ for the distribution of the issuer's asset value $Z = X + Y$ may force $Z$ 
to have a small variance. As a consequence, it may happen that model \eqref{eq:lognormal} is not able 
to fit the given risk characteristics. But recall the comments from Remark~\ref{rm:consequence} regarding
the appropriate choice of time horizons for the risk characteristics. To some extent, with the right
time horizons, the problem of not having an exactly fitting two-assets model thus might be mitigated.

Due to the complex structure of the equation system \eqref{eq:2optim1} and \eqref{eq:2optim2} it 
would be hard to try and analyse the particulars of the question when exactly there is no solution to the 
system. To better illustrate the cause of the problem, in the rest of the section we instead analyse a toy model based on
two jointly normal distributed asset value variables.

\subsection{Model existence in the case of two jointly normal asset values}
\label{se:2normal}

In this example, we leave all assumptions from Sections~\ref{se:2assetsEL} and \ref{se:2lognormal} unchanged 
but replace \eqref{eq:lognormal} with
\begin{equation}\label{eq:normal}
X \ = \ \mu + \sigma\,\xi, \qquad Y \ = \ \nu + \tau\,\eta,
\end{equation}
where we have $\mu, \nu \in \mathbb{R}$, $\sigma, \tau > 0$ and $(\xi, \eta) \sim \mathcal{N}\left(
\begin{pmatrix}
0 \\ 0
\end{pmatrix}, \begin{pmatrix}
1 & \varrho \\
\varrho & 1
\end{pmatrix}\right)$ for some $\varrho \in [0,1]$. This is not a satisfactory assumption for modelling
asset values because $X$ and $Y$ may take negative values. But it is helpful for the purpose of illustrating
potential calibration problems because the sum of two jointly normal random variables is again normal. 
As a consequence, in this case, in contrast to the lognormal case of Equations \eqref{eq:2optim1}
and \eqref{eq:2optim2} (or \eqref{eq:2com1}
and \eqref{eq:2com2}) we obtain a closed-form solution for the calibration equations.

Under \eqref{eq:normal} the equations for the parameters $\mu$ and $\sigma$ are given by (cf.\ Section~\ref{se:ESLocSc})
\begin{equation}\label{eq:coverNormal}
\begin{split}
\mu & \ = \ (1+v)\,C - \sigma\,\Phi^{-1}(p_{\mathrm{cover}}), \\
\sigma & \ = \  \frac{(1+v)\,C\,EL_{\mathrm{cover}}}
	{p_{\mathrm{cover}}\,\Phi^{-1}(p_{\mathrm{cover}})+\varphi\bigl(\Phi^{-1}(p_{\mathrm{cover}})\bigr)}.
\end{split}
\end{equation} 
The equation system \eqref{eq:2optim1} and \eqref{eq:2optim2} for $\nu$ and $\tau$ is replaced by
\begin{equation}\label{eq:otherNormal}
\begin{split}
\nu & \ = \ D_{\mathrm{issuer}} - \mu -  \psi\,\Phi^{-1}(p_{\mathrm{issuer}}), \\
\tau & \ = \ \sqrt{\psi^2 - (1-\varrho^2)\,\sigma^2} - \sigma\,\varrho, \\
\psi & \ = \  \frac{D_{\mathrm{issuer}}\,p_{\mathrm{issuer}}\,LGD_{\mathrm{issuer}}}{p_{\mathrm{issuer}}\,\Phi^{-1}(p_{\mathrm{issuer}})+
	\varphi\bigl(\Phi^{-1}(p_{\mathrm{issuer}})\bigr)}.
\end{split}
\end{equation}
With regard to \eqref{eq:otherNormal}, it is easy to see that
\begin{equation}\label{eq:solvability}
\tau > 0 \qquad \iff \qquad \psi > \sigma.
\end{equation} 
The parameter $\psi$ is actually the implied standard deviation of the bond issuer's total asset value. 
As observed above, condition \eqref{eq:solvability} is natural for the existence
of a two-assets model with PD and LGD constraints for the cover pool and the issuer's entire portfolio.
However, while in the case of assumption \eqref{eq:normal} the condition is indeed necessary and sufficient 
for the existence of a model based on a joint normal distribution, it is unclear whether a similar 
statement holds true in the case of assumption \eqref{eq:lognormal}. Note that even if \eqref{eq:solvability}
is satisfied the resulting model still might not make economic sense because it could occur that $\nu \le 0$.

\section{The lognormal one-asset model}
\label{se:1lognormal}

As shown in Section~\ref{se:2lognormal}, it may be impossible to exactly fit a two-assets model for the
covered bonds calculations. Therefore, in this section we consider how
to implement a lognormal one-asset model and 
explore how to make it sensitive with regard to differences in the riskiness
of cover pool and remaining assets.

\paragraph{Lognormal one-asset model.} The uncertain future value $A$ of the total assets 
of the covered bonds issuer 
(including the cover pool assets) is given by
\begin{equation}\label{eq:OneAsset}
A\ =\ \exp(\kappa + \psi\,\xi),
\end{equation}
where $\xi$ is a standard normal random variable. The value of the cover pool assets is
$\varepsilon\,A$ for a constant $\varepsilon \in (0,1]$. The number $\varepsilon$ is 
called \emph{asset encumbrance ratio}.

In this section, basically we again assume the setting of Section~\ref{se:2assets}, 
with $X = \varepsilon\,A$ and $Y = (1-\varepsilon)\,A$:
\begin{itemize} 
\item  The calibration of the parameters $\kappa$ and $\psi$ of 
the lognormal one-asset model then is straightforward (see the comments on the derivation of
\eqref{eq:coverpool}) if the issuer's PD
$p_{\mathrm{issuer}}$ and LGD $LGD_{\mathrm{issuer}}$ are known:
\begin{equation}\label{eq:1calibration}
\begin{split}
\kappa & \ = \ \log(C+S+U) - \psi\,\Phi^{-1}(p_{\mathrm{issuer}}),\\
0 & \ = \ \Phi\bigl(\Phi^{-1}(p_{\mathrm{issuer}})-\psi\bigr) -
		 \bigl(p_{\mathrm{issuer}}-EL_{\mathrm{issuer}}\bigr)\,
		 \exp\bigl(\psi\,\Phi^{-1}(p_{\mathrm{iusser}})-\psi^2/2\big),
\end{split}
\end{equation}
with $EL_{\mathrm{issuer}} = p_{\mathrm{issuer}}\,LGD_{\mathrm{issuer}}$. By 
Proposition~\ref{pr:lognormal} the equation system \eqref{eq:1calibration} has a unique solution
$(\kappa, \psi)$.
\item In principle, the encumbrance ratio $\varepsilon$ could be determined as
\begin{equation}\label{eq:alterEps}
\varepsilon\ =\ (1+v)\,\frac{C}{A_0},
\end{equation} 
where $A_0 > C + S + U$ are the issuer's total
assets at time 0 (today), $v$ is the level of over-collateralisation of the covered bonds and $C$
is the face value of the covered bonds. 
\end{itemize}
In practice, the value of $A_0$ might not be accurately known, for a number of possible reasons. 
In particular, under an amortised cost accounting regime the sum of the asset values might
be considered a too unreliable and potentially too conservative estimate of their true total value.
That is why, in this section, we explore how to infer $\varepsilon$ 
implicitly from the level of over-collateralisation $v$ and the expected loss of the cover pool.

\subsection{Adjusting the asset encumbrance ratio}
\label{se:adjusting}
 
Assume that in addition to the debt amounts $C$, $S$, and $U$ and level of over-collateralisation $v$
as in Section~\ref{se:2assets} and parameters $\kappa$, $\psi$ (calibrated according to 
\eqref{eq:1calibration}) a value $EL_{\mathrm{cover}}$ of expected loss for the cover pool is given.

The idea is to find $\varepsilon$ such that 
\begin{equation}\label{eq:adjust}
(1+v)\,C\,EL_{\mathrm{cover}} \ = \ \mathrm{E}\bigl[\bigl((1+v)\,C - \varepsilon\,A\bigr)
	\mathbf{1}_{\{\varepsilon\,A < (1+v)\,C\}}\bigr].
\end{equation}
Actually, \eqref{eq:adjust} is equivalent to the second equation of \eqref{eq:coverpool}.
Recall that we represent the value of the portfolio as $A = e^{\kappa+\psi\,\xi}$ with a standard normal 
random variable $\xi$. With $\vartheta = \kappa + \log(\varepsilon)$ therefore we may rewrite \eqref{eq:adjust} as
\begin{align}\label{eq:adjust2}
(1+v)\,C\,EL_{\mathrm{cover}} & \ = \ \mathrm{E}\bigl[\bigl((1+v)\,C - e^{\vartheta+\psi\,\xi}\bigr)
	\mathbf{1}_{\{e^{\vartheta+\psi\,\xi} < (1+v)\,C\}}\bigr]\notag\\
& \ = \ (1+v)\,C\,\Phi\left(\frac{\log\bigl((1+v)\,C\bigr)-\vartheta}{\psi}\right)
	- e^{\vartheta+\psi^2/2}\,\Phi\left(\frac{\log\bigl((1+v)\,C\bigr)-\vartheta}{\psi} - \psi\right)\notag\\
& \ \stackrel{\text{def}}{=}\ f(\vartheta). 
\end{align}
Based on \eqref{eq:adjust2}, it can be shown that \eqref{eq:adjust} always has a unique solution 
$\varepsilon$:
\begin{proposition}\label{pr:unique}
 	There is exactly one $\vartheta \in \mathbb{R}$ such that $f(\vartheta) = (1+v)\,C\,EL_{\mathrm{cover}}$, with 
 	$f(\vartheta)$ defined by \eqref{eq:adjust2}.
\end{proposition}
\textbf{Proof.} Obviously, we have $\lim_{\vartheta \to -\infty} f(\vartheta) = (1+v)\,C$. The limit behaviour 
$\lim_{\vartheta \to \infty} f(\vartheta) = 0$ is less obvious but follows from \eqref{eq:ineq}. 
Since $0 < EL_{\mathrm{cover}} < 1$, therefore by continuity of
$f$ there is a solution $\vartheta$ to $f(\vartheta) = (1+v)\,C\,EL_{\mathrm{cover}}$. Since 
$\vartheta_1 < \vartheta_2$ implies $\bigl\{e^{\vartheta_2+\psi\,\xi} < (1+v)\,C\bigr\} \subset 
\bigl\{e^{\vartheta_1+\psi\,\xi} < (1+v)\,C\bigr\}$ we obtain for $\vartheta_1 < \vartheta_2$
\begin{align*}
	\mathrm{E}\bigl[\bigl((1+v)\,C - e^{\vartheta_1+\psi\,\xi}\bigr)
	\mathbf{1}_{\{e^{\vartheta_1+\psi\,\xi} < (1+v)\,C\}}\bigr] & \ \ge \
	\mathrm{E}\bigl[\bigl((1+v)\,C - e^{\vartheta_1+\psi\,\xi}\bigr)
	\mathbf{1}_{\{e^{\vartheta_2+\psi\,\xi} < (1+v)\,C\}}\bigr] \\
	&\ >\ \mathrm{E}\bigl[\bigl((1+v)\,C - e^{\vartheta_2+\psi\,\xi}\bigr)
	\mathbf{1}_{\{e^{\vartheta_2+\psi\,\xi} < (1+v)\,C\}}\bigr].
\end{align*}
This implies uniqueness of the solution of $f(\vartheta) = (1+v)\,C\,EL_{\mathrm{cover}}$. \hfill \qed

By Proposition~\ref{pr:unique}, we can define a lognormal one-asset model for covered bonds expected loss
in the following way:
\begin{definition}\label{de:one-asset}
Assume that the issuer's PD $p_{\mathrm{issuer}}$, the issuer's expected loss $EL_{\mathrm{issuer}}$, 
the cover pool expected loss $EL_{\mathrm{cover}}$, 
the debt face values $C$ for the covered bonds, $S$ for the senior unsecured debt, 
$U$ for the junior unsecured debt, and the level of over-collateralisation $v$ of the covered bonds are 
given. The \emph{adjusted lognormal one-asset model} then is specified by the following two 
calibration steps:
\begin{itemize}
\item Parametrise the asset value $A = e^{\kappa + \psi\,\xi}$ with $\xi \sim \mathcal{N}(0,1)$, $\kappa \in 
\mathbb{R}$ and $\psi > 0$ by solving (for $\kappa$ and $\psi$) the equation system \eqref{eq:1calibration}.
\item Determine the \emph{adjusted encumbrance ratio} $\varepsilon$ by solving \eqref{eq:adjust2} for 
the unique solution $\vartheta$ and setting $\varepsilon = \min(1, e^{\vartheta - \kappa})$.
\end{itemize} 
\end{definition}
The 100\% cap of $\varepsilon$ in the second calibration step of Definition~\ref{de:one-asset} is needed 
because while $\varepsilon$ is always positive there is no guarantee that it is smaller than 1, in 
particular if the values  of $EL_{\mathrm{cover}}$ and $EL_{\mathrm{issuer}}$ are very different. 
A value of $\varepsilon$ greater than 100\%
would not make economic sense because then the value of the non-cover-pool assets in the issuer's 
portfolio would be negative. Actually, in a way similar to Proposition~\ref{pr:syssolution}, we can 
state which values can be exactly realised for $EL_{\mathrm{cover}}$ for given 
characteristics of the issuer's portfolio . 
The following observation is a consequence of the fact that the function $f(\vartheta)$ in
\eqref{eq:adjust2} is a decreasing function of the parameter $\vartheta$ (as shown in the proof of
Proposition~\ref{pr:unique}).
\begin{remark}
In the setting of Definition~\ref{de:one-asset}, the minimum expected loss of the cover pool that can be 
matched with the adjusted lognormal one-asset model is
$$EL_{\mathrm{cover}, \min} \ = \  
\Phi\left(\frac{\log\bigl((1+v)\,C\bigr)-\kappa}{\psi}\right)
	- \frac{e^{\kappa+\psi^2/2}}{(1+v)\,C}\,
	\Phi\left(\frac{\log\bigl((1+v)\,C\bigr)-\kappa}{\psi} - \psi\right).$$
\end{remark}
How different is the adjusted lognormal one-asset model from the lognormal two-assets model as discussed 
in Sections~\ref{se:2assetsEL} and \ref{se:CaseLognormal}? The following proposition is useful for
figuring this out.
\begin{proposition}\label{pr:equivalent}
Consider the following two bivariate random vectors:
\begin{itemize}
\item $(\varepsilon\,A,\, (1-\varepsilon)\,A)$ with $0 < \varepsilon < 1$, 
$A = e^{\kappa + \psi\,\zeta}$, $\zeta \sim \mathcal{N}(0,1)$, $\kappa \in \mathbb{R}$, $\psi > 0$. Call
this \emph{Model 1}.
\item $(X, Y)$ with $X = e^{\mu+\sigma\,\xi}$, $Y = e^{\nu + \tau\,\eta}$, $\begin{pmatrix} \xi \\ \eta
	\end{pmatrix} \sim \mathcal{N}\left(\begin{pmatrix} 0 \\ 0
	\end{pmatrix}, \begin{pmatrix} 1 & \varrho \\ \varrho & 1
	\end{pmatrix}\right)$, $\mu, \nu \in \mathbb{R}$, $\sigma, \tau > 0$, $\varrho \in [-1,1]$. Call this
	\emph{Model 2}.
\end{itemize}
Model~1 and Model~2 are equivalent in the sense that the respective random vectors have the same 
distributions if and only if 
\begin{equation}\label{eq:equiv}
\psi = \tau = \sigma,\quad \varrho =1,\quad \varepsilon = \frac{e^\mu}{e^\mu+e^\nu},\quad
\kappa = \log(e^\mu+e^\nu).
\end{equation}
\end{proposition}
\textbf{Proof.} Assume first that \eqref{eq:equiv} is satisfied. By $\varrho =1$ and $\sigma=\tau$ 
we then have almost surely that
$$(X,Y) \ =\ \bigl(e^{\mu+\sigma\,\xi}, \,e^{\nu+\sigma\,\xi} \bigr).$$
Making use of $\psi = \sigma$ and of the representations of $\varepsilon$ and $\kappa$ in \eqref{eq:equiv} we obtain
$$e^{\mu+\sigma\,\xi} = \varepsilon\,e^{\kappa+\psi\,\xi}, \quad e^{\nu+\sigma\,\xi} = (1-\varepsilon)\,e^{\kappa+\psi\,\xi}.$$
Since $\xi \sim \mathcal{N}(0,1)$, this implies the 'if' part of Proposition~\ref{pr:equivalent}. 

Assume 
now that the bivariate distributions implied by Model~1 and Model~2 respectively are equal.
Then in particular, the distributions of $\log(\varepsilon\,A) = \log(\epsilon) + \kappa+ 
\psi\,\zeta$ and $\log(X) = \mu + \sigma\,\xi$ are equal. Since both $\zeta$ and $\xi$ are standard normal,
it follows that $\mu = \log(\epsilon) + \kappa$ and $\sigma = \psi$. The same rationale applied to 
$(1-\varepsilon)\,A$ and $Y$ gives $\nu = \log(1-\epsilon) + \kappa$ and $\tau = \psi$. Hence we have
$\psi = \tau = \sigma$ and $\mu - \log(\epsilon) = \nu - \log(1-\epsilon)$ which gives
$\varepsilon = \frac{e^\mu}{e^\mu+e^\nu}$. This implies $\kappa = \mu - \log(\epsilon) = 
\log(e^\mu+e^\nu)$. By $\psi > 0$ the components of $(\varepsilon\,A,\, (1-\varepsilon)\,A)$ are comonotonic.
Hence also $X$ and $Y$ are comonotonic which implies that $\log(X) = \mu + \sigma\,\xi$ and 
$\log(Y)= \nu + \tau\,\eta$ are comonotonic. However, two jointly normal random variables are only
comonotonic if their correlation is 1. This completes the proof of 'only if'. \hfill\qed

By Proposition~\ref{pr:equivalent}, for homogeneous portfolios, 
the adjusted lognormal one-asset model of Definition~\ref{de:one-asset}
is a comonotonic special case of the lognormal two-assets model from Section~\ref{se:2assetsEL}.
To formally show this, we consider the following type of lognormal two-assets model.

\begin{definition}\label{de:2lognormals}
A lognormal two-assets model as specified by \eqref{eq:lognormal} is called \emph{margins-calibrated} 
if it is calibrated in the following way:
\begin{itemize}
\item Debt face values $C$ for covered bonds, $S$ for senior unsecured debt and $U$ for junior debt are 
given.
\item A level $v$ of over-collateralisation is given.
\item A correlation parameter $\varrho$ is chosen on the basis of expert judgment or a separate calibration
exercise.
\item Parameters $\mu$ and $\sigma$ are determined by solving the equation system \eqref{eq:coverpool}.
\item Parameters $\nu$ and $\tau$ are determined by solving the equation system
\begin{equation}\label{eq:otherpool}
\begin{split}
\nu & \ = \ \log(S+U-v\,C) - \tau\,\Phi^{-1}(p_{\mathrm{other}}),\\
0 & \ = \ \Phi\bigl(\Phi^{-1}(p_{\mathrm{other}})-\tau\bigr) -
		 \bigl(p_{\mathrm{other}}-EL_{\mathrm{other}}\bigr)\,
		 \exp\bigl(\tau\,\Phi^{-1}(p_{\mathrm{other}})-\tau^2/2\big),
\end{split}
\end{equation}
where $p_{\mathrm{other}}$ and $EL_{\mathrm{other}}$ denote the stand-alone probability of loss and the
expected loss respectively of the issuer's portfolio without the cover pool. 
\end{itemize}
\end{definition}
The term $-v\,C$ is a consequence of having the term $1+v$ in \eqref{eq:coverpool} to reflect the 
over-collateralisation of the cover pool. The term is a correction needed to make sure that the total
debt is $C+S+U$.
The calibration approach in Definition~\ref{de:2lognormals} is computationally less demanding 
than the calibration approach from Section~\ref{se:CaseLognormal} and always
feasible. But as mentioned in
Section~\ref{se:CaseLognormal}, often no estimates of $p_{\mathrm{other}}$ and 
$EL_{\mathrm{other}}$ might be available as input to \eqref{eq:otherpool}. Nonetheless, for comparison with
the adjusted lognormal one-asset model and illustrative purposes, the margins-calibrated 
two-assets model is helpful as demonstrated in the following remark and Section~\ref{se:numerical} 
respectively.
\begin{remark}\label{rm:equiv}
In the comonotonic case ($\varrho = 1$) of Definition~\ref{de:2lognormals}, 
if in addition $p_{\mathrm{other}} = p_{\mathrm{cover}}$
and $EL_{\mathrm{other}} = EL_{\mathrm{cover}}$ (i.e.\ the issuer's portfolio is homogeneous in terms
of risk) then
by Proposition~\ref{pr:lognormal} if follows that $\sigma = \tau$. Hence by Proposition~\ref{pr:equivalent}
the two-assets model can be represented as a lognormal one-asset model. This one-asset model actually is an 
adjusted lognormal one-asset model in the sense of Definition~\ref{de:one-asset} with
$p_{\mathrm{issuer}} = p_{\mathrm{cover}}$ and $EL_{\mathrm{issuer}} = EL_{\mathrm{cover}}$. 
This follows from the observations that 
\begin{itemize}
\item with $\varepsilon  =   \frac{(1+v)\,C}{C + S + U}$ the calibration equation 
systems \eqref{eq:1calibration} for the one-asset model and \eqref{eq:coverpool} and \eqref{eq:otherpool}
for the two-assets model are all equivalent,
\item by \eqref{eq:equiv} we have $e^\mu = \varepsilon\,e^\kappa$, and
\item \eqref{eq:coverpool} implies \eqref{eq:adjust} with 
$\varepsilon\,A$ replaced by $e^{\mu+\sigma\,\xi}$ where $\xi$ is standard normal.   
\end{itemize}
\end{remark}

\subsection{Expected loss for the adjusted lognormal one-asset model}
\label{se:1formulae} 

In the case of a lognormal one-asset model like \eqref{eq:OneAsset}, with encumbrance ratio $\varepsilon$
calibrated according to \eqref{eq:alterEps} or as in Definition~\ref{de:one-asset}, it is 
straightforward to derive the
expected loss formulae for the loss variables defined by \eqref{eq:ev1},  \eqref{eq:ev2} and
\eqref{eq:ev3} (with $X= \varepsilon\,A$ and $Y=(1-\varepsilon)\,A$). For the readers' convenience,
we list these formulae in the following.
\begin{itemize}
\item Probabilities of the loss events:
\begin{subequations}
\begin{align}
\mathrm{P}[A < C+S+U,\, A \ge C+S] & = 
\Phi\left(\frac{\log(C+S+U))-\mu}{\sigma}\right) -
\Phi\left(\frac{\log(C+S)-\mu}{\sigma}\right),\\
\mathrm{P}[A < C+S,\, \varepsilon\,A \ge C] & = 
	\Phi\left(\frac{\log\bigl(\max(C/\varepsilon, C+S)\bigr)-\mu}{\sigma}\right) -
\Phi\left(\frac{\log(C/\varepsilon)-\mu}{\sigma}\right),\\
\mathrm{P}[A < C+S,\, \varepsilon\,A < C] & =
\Phi\left(\frac{\log\bigl(\min(C/\varepsilon, C+S)\bigr)-\mu}{\sigma}\right).
\end{align}
\end{subequations}
\item Expected loss for the different types of debt (covered bonds, senior unsecured and junior unsecured):
\begin{subequations}
\begin{align}
C\,\mathrm{E}[L_C] & \ =\ \int\limits^{\frac{\log(\min(C/\varepsilon, C+S))-\mu}{\sigma}}_{-\infty}
 \varphi(x)\,\frac{\bigl(C-\varepsilon\,e^{\mu+\sigma\,x}\bigr)
 	\bigl(C+S-e^{\mu+\sigma\,x}\bigr)}{C+S-\varepsilon\,e^{\mu+\sigma\,x}}\,dx,\\
S\,\mathrm{E}[L_S] & \ =\ S \int\limits^{\frac{\log(\min(C/\varepsilon, C+S))-\mu}{\sigma}}_{-\infty}
 \varphi(x)\,\frac{C+S-e^{\mu+\sigma\,x}}{C+S-\varepsilon\,e^{\mu+\sigma\,x}}\,dx 
 + (C+S)\,\mathrm{P}[A < C+S,\, \varepsilon\,A \ge C]\notag\\
& \qquad - e^{\mu+\sigma^2/2} 
	\left\{\Phi\left(\frac{\log\bigl(\max(C/\varepsilon, C+S)\bigr)-\mu}{\sigma}-\sigma\right) -
\Phi\left(\frac{\log(C/\varepsilon)-\mu}{\sigma}-\sigma\right)\right\},\\
U\,\mathrm{E}[L_U] & \ =\ U\,\mathrm{P}[A < C+S] + (C+S+U)\,\mathrm{P}[A < C+S+U,\, A \ge C+S]\notag\\
& \qquad - e^{\mu+\sigma^2/2} 
	\left\{\Phi\left(\frac{\log(C+S+U)-\mu}{\sigma}-\sigma\right) -
\Phi\left(\frac{\log(C+S)-\mu}{\sigma}-\sigma\right)\right\}.
\end{align} 
\end{subequations}
\end{itemize}

\section{Numerical examples}
\label{se:numerical}

Due to the total lack of credit loss data related to covered bonds, we cannot directly test the models 
described in Sections~\ref{se:2assetsEL}, \ref{se:CaseLognormal} and \ref{se:adjusting}. What we can do is checking model results for
economic plausibility and comparing the margins-calibrated two-assets and adjusted one-asset models. 
Hence, in this section, we 
present some results to illustrate how the models work in the following three scenarios:
\begin{itemize}
\item Increasing correlation between the values of the cover pool assets and the issuer's other assets
respectively.
\item Increasing asset encumbrance by increasing the proportion of the total debt consisting of covered bonds.
\item Using the adjusted lognormal one-asset model instead of the lognormal two-assets model.
\end{itemize}

\subsection{Impact of dependence between cover pool and remaining assets} 
\label{se:dependence}

We use the margins-calibrated 
lognormal two-assets model as described in Definition~\ref{de:2lognormals} to illustrate the impact of
correlation on the expected loss of covered bonds (senior secured debt), senior unsecured debt and junior 
(unsecured) debt respectively. The input parameters are as follows:
\begin{itemize}
\item $C=0.3$, $S=0.6$, $U=0.1$, $v=0.2$.
\item $p_{\mathrm{cover}} = p_{\mathrm{other}} = 0.01$, 
$LGD_{\mathrm{cover}} = LGD_{\mathrm{other}} = 0.45$.
\item $\varrho \in \{0, 0.3, 0.6, 0.9, 1.0\}$. This generates a spectrum of dependence increasing
from independence ($\varrho = 0$) to comonotonicity ($\varrho = 1$).
\end{itemize}
The CGFS recently found in a survey of 60 European banks 28.5\% median asset encumbrance in the banks'
balance sheets \citep[][Graph~3]{CGFS49Paper}. This finding was the motivation for the choice of the debt
amounts in the debt hierarchy of this example, together with the consideration that the proportion of junior
debt in general is small compared to the proportion of senior debt.

The results of the calculation are shown in Table~\ref{tab:1}. The impact of increasing correlation
on the expected loss is very strong across all debt types but strongest for the covered bonds (senior secured
debt). This behaviour is intuitive because strong dependence implies a larger likelihood of a big part of 
the issuer's assets being erased. In the following two examples, 
we consider the comonotonic case ($\varrho = 1$) only, 
thus adopting a worst case perspective that, in particular, might be appropriate for stress testing
purposes. 
\begin{table}
\begin{center}
\caption{\small{}Illustration of expected loss (as \% of exposure) for covered bonds, senior unsecured debt and junior 
debt
as function of the correlation (as \%) between the cover pool logarithmic asset value and the logarithm of the 
value of the remaining debt.
Approach and parameters are explained in Section~\ref{se:dependence}.}\label{tab:1}
\begin{tabular}{|l|r|r|r|r|r|}
\hline
Correlation $\varrho$&$0.0$&$30.0$&$60.0$&$90.0$&$100.0$ \\ \hline \hline
Covered bonds EL&$0.002$&$0.014$&$0.057$&$0.173$&$0.257$ \\ \hline
Senior unsecured EL&$0.007$&$0.039$&$0.132$&$0.345$&$0.465$ \\ \hline
Junior EL &$0.020$&$0.097$&$0.294$&$0.711$&$0.943$ \\ \hline
All EL&$0.007$&$0.038$&$0.126$&$0.330$&$0.450$ \\ \hline
\end{tabular}
\end{center} 
\end{table}
 
\subsection{Impact of asset encumbrance}
\label{se:encumbrance}

Here we use the adjusted 
lognormal one-asset model as described in Definition~\ref{de:one-asset} to illustrate the impact of
asset encumbrance on the expected loss of the senior unsecured debt and the junior 
(unsecured) debt respectively. The input parameters are as follows:
\begin{itemize}
\item $p_{\mathrm{issuer}} = 0.01$, $LGD_{\mathrm{issuer}} = 0.45$. 
\item $EL_{\mathrm{cover}} =  0.0045$.
\item $C \in \{0, 0.1, 0.2, \ldots, 0.8\}$, $S=0.9-C$, $U=0.1$, $v=0.2$. This generates a spectrum of 
asset encumbrance increasing
from none ($C = 0$) to strong ($C = 0.8$).
\end{itemize}
Note that according to Remark~\ref{rm:equiv}, this specific adjusted one-asset model 
(with $EL_{\mathrm{issuer}} = EL_{\mathrm{cover}}$) is equivalent to 
the margins-calibrated model with $p_{\mathrm{cover}} = p_{\mathrm{other}}$ and 
$LGD_{\mathrm{cover}} = LGD_{\mathrm{other}}$. Moreover, the adjusted asset encumbrance ratio in this
case is $\varepsilon  =  \frac{(1+v)\,C}{C + S + U}$.

The results of the calculation are shown in Table~\ref{tab:2}. Due to the subordination of the junior debt
when it comes to distributing the recoveries, the expected loss for the junior debt does not depend on
the proportions of secured and unsecured senior debt. There is, however, a significant dependence of
the senior unsecured expected loss on the asset encumbrance ratio. The dependence becomes the stronger the
more the asset encumbrance ratio approaches 100\%. But also the per unit expected loss of the
covered bonds increases with increasing asset encumbrance ratio. This is caused by the fact that the
buffer for the covered bonds provided by lower ranking and unsecured debt shrinks with the increasing 
proportion of the  total debt consisting of the covered bonds. Note that nonetheless 
in Table~\ref{tab:2} irrespective 
of the changes in the covered bonds exposure the expected loss of the issuer's entire portfolio remains 
constant at 0.45\%.

\begin{table}
\begin{center}
\caption{\small{}Illustration of expected loss (as \% of exposure) for covered bonds, 
senior unsecured debt and junior debt
as function of the covered bonds exposure (as \% of total debt). 
Approach and parameters are explained in Section~\ref{se:encumbrance}.}\label{tab:2}
\begin{tabular}{|l|r|r|r|r|r|r|r|r|r|}
\hline
Covered bonds exposure $C$&$0.0$&$10.0$&$20.0$&$30.0$&$40.0$&$50.0$&$60.0$&$70.0$&$80.0$ \\ \hline\hline
Adjusted encumbrance ratio&$0.0$&$12.0$&$24.0$&$36.0$&$48.0$&$60.0$&$72.0$&$84.0$&$96.0$ \\ \hline
Covered bonds EL& NA&$0.237$&$0.246$&$0.257$&$0.269$&$0.283$&$0.300$&$0.320$&$0.347$ \\ \hline
Senior unsecured EL&$0.395$&$0.415$&$0.438$&$0.465$&$0.496$&$0.536$&$0.587$&$0.658$&$0.777$ \\ \hline
Junior EL&$0.943$&$0.943$&$0.943$&$0.943$&$0.943$&$0.943$&$0.943$&$0.943$&$0.943$ \\ \hline
\end{tabular}
\end{center} 
\end{table}

\subsection{Lognormal two-assets vs adjusted lognormal one-asset}
\label{se:vs}

Remark~\ref{rm:equiv} presents sufficient conditions to coincide for the margins-calibrated lognormal two-assets
model as in Definition~\ref{de:2lognormals} and the adjusted lognormal one-asset model as in 
Definition~\ref{de:one-asset}. In this section, we show that there is no longer equivalence 
if the assumption $p_{\mathrm{other}} = p_{\mathrm{cover}}$
and $EL_{\mathrm{other}} = EL_{\mathrm{cover}}$ is violated, i.e.\ if the cover pool and the pool
of the issuer's remaining assets are no longer homogeneous in terms of risk.

With regard to the debt exposures and the level of over-collateralisation, we use the same input
parameters as in Section~\ref{se:dependence}, i.e.
$$C=0.3,\quad S=0.6,\quad U=0.1,\quad v=0.2.$$
For the two-assets model according to Definition~\ref{de:2lognormals}, we use use the following
parameters:
\begin{itemize}
\item $\varrho = 1$, i.e.\ we consider the comonotonic case.
\item $p_{\mathrm{other}} = 0.01$, $LGD_{\mathrm{other}} = 0.45$.
\item We keep $EL_{\mathrm{cover}}$ fix at $0.003$, but choose $LGD_{\mathrm{cover}} \in
\{0.3, 0.45, 0.6\}$ and determine $p_{\mathrm{cover}}$ by $p_{\mathrm{cover}} = 
\frac{EL_{\mathrm{cover}}}{LGD_{\mathrm{cover}}}$. 
\end{itemize}
The corresponding one-asset model according to Definition~\ref{de:one-asset} is calibrated 
as follows:
\begin{itemize}
\item We determine the parameters $\mu$, $\sigma$, $\nu$ and $\tau$ for the two-assets model as
described in Definition~\ref{de:2lognormals}. 
\item We then compute $p_{\mathrm{issuer}}$ as the sum of the probabilities given by \eqref{eq:p1}, \eqref{eq:p2} and \eqref{eq:p3}. Similarly, we compute 
$EL_{\mathrm{issuer}}$ as the sum of the right-hand sides of \eqref{eq:com1},  \eqref{eq:com2}
and \eqref{eq:com3}, divided by $C+S+U$.
\item As input parameters for the calibration of the one-asset model as described in 
Definition~\ref{de:one-asset} we use $p_{\mathrm{issuer}}$ and $EL_{\mathrm{issuer}}$ to 
determine $\kappa$ and $\psi$, and $EL_{\mathrm{cover}}$ and $v$ to determine $\varepsilon$.
\end{itemize}
Table~\ref{tab:3} shows that as long the risk characteristics of the issuer's entire portfolio 
are identical (in terms of PD and expected loss) as input to the two models 
and the difference in expected loss of the cover pool and the pool of the remaining assets is not
too big,
the resulting expected loss values 
for the different debt types according to the two-assets model and the one-asset model are quite
close. Material differences in the results by the two models may be incurred especially 
if the cap of the asset
encumbrance ratio as suggested in Definition~\ref{de:one-asset} kicks in. This may happen 
if the expected loss of the issuer's entire portfolio is much larger than the expected loss of the
cover pool. Then it is possible that
the adjusted one-asset model outputs an expected loss value for the covered bonds which is 
larger than the expected loss of the cover pool -- a quite unintuitive result. 

Another interesting observation from Table~\ref{tab:3} is that the shape of the cover pool
asset value distribution (as determined by its probability of loss and loss-given-default rate) 
has some impact on both the portfolio-wide expected loss and the expected losses of the debt types.
In particular, the more skewed the cover pool asset value distribution becomes, the larger the covered bonds
expected loss grows.   

\begin{table}
\begin{center}
\caption{\small{}Results of the margins-calibrated lognormal two-assets model compared with
results of the adjusted lognormal one-asset model. For model calibration and input parameters
see Section~\ref{se:vs}. All numbers are percentages.}\label{tab:3}
\begin{tabular}{|l||r|r||r|r||r|r|}\hline
$LGD_{\mathrm{cover}}$ & \multicolumn{2}{|c|}{30} & \multicolumn{2}{|c|}{45} &
	\multicolumn{2}{|c|}{60} \\ \hline
$p_{\mathrm{cover}}$ & \multicolumn{2}{|c|}{1.000} & \multicolumn{2}{|c|}{0.667} &
	\multicolumn{2}{|c|}{0.500} \\ \hline \hline
$p_{\mathrm{issuer}}$ & \multicolumn{2}{|c|}{1.000} & \multicolumn{2}{|c|}{0.848} &
	\multicolumn{2}{|c|}{0.655} \\ \hline
$EL_{\mathrm{issuer}}$ & \multicolumn{2}{|c|}{0.396} & \multicolumn{2}{|c|}{0.382} &
	\multicolumn{2}{|c|}{0.348} \\ \hline \hline
	& 2 assets & 1 asset & 2 assets & 1 asset & 2 assets & 1 asset \\ \hline
Covered bonds EL&$0.147$&$0.168$&$0.189$&$0.188$&$0.216$&$0.211$ \\ \hline
Senior unsecured EL&$0.431$&$0.421$&$0.409$&$0.409$&$0.368$&$0.371$ \\ \hline
Junior EL&$0.931$&$0.929$&$0.799$&$0.799$&$0.629$&$0.628$ \\ \hline
\end{tabular}
\end{center} 
\end{table}

\section{Conclusions}
\label{se:conclusions}

Due to the growing importance of covered bonds as a funding tool for banks and as investment for risk-averse investors,
quantitative approaches to expected loss for covered bonds and the impact of asset encumbrance by covered bonds
on expected loss for senior unsecured debt are of high interest for practitioners and researchers.

In this paper, we have studied a natural lognormal two-assets model for the expected loss of covered bonds. It has 
turned out, however, that in some circumstances exact calibration of this model is impossible. Therefore,
we have suggested a simple and computationally efficient adjusted lognormal one-asset approach 
to these issues. Input data requirements by this approach are light: PD and LGD of the 
covered bonds issuer, expected
loss of the cover pool as well as the distribution of debt types in the issuer's portfolio and the level of
over-collateralisation of the covered bonds. The model therefore
can be applied in situations like stress testing 
where little information is available and many bonds have to be evaluated in short time.

We have demonstrated that for homogeneous portfolios the one-asset approach is a special case of the more natural two-assets 
approach whose calibration, however, is difficult and sometimes not feasible at all. 
Backtesting of the two models discussed is impossible because
never any covered bonds defaults were observed. Nonetheless, we have presented numerical examples 
that suggest that the
results by the proposed 'adjusted lognormal one-asset model' concur with intuition as long as the
risk levels of the issuer's entire portfolio and of the cover pool of the covered bonds are not materially 
different.


\bibliographystyle{plainnat}

\addcontentsline{toc}{section}{References}

\appendix

\section{Appendix: Mean-variance matching}
\label{se:MeanVar}

Fitting a two-parameter distribution to given values of mean and variance is a common approach in finance, 
in particular when little is known about the target distribution otherwise. This approach is referred to
as \emph{mean-variance matching} and is a special case of \emph{moment matching} or the \emph{method of moments}.

We start the description of mean-variance matching by revisiting the location-scale family associated 
with some random variable $X$ 
in the sense of Definition~\ref{de:LocScale}.
Obviously, if $X$ has a positive and finite variance, for each pre-defined pair $(\mu, \sigma) \in \mathbb{R} \times (0, \infty)$
there is exactly one distribution in the location-scale distribution family of $X$ with mean $\mu$ and standard deviation 
$\sigma$. To identify the parameters $m$ and $s$ of this distribution, we have to solve the equation system
\begin{subequations}
\begin{equation}
\begin{split}
    \mu &\ =\ \mathrm{E}[m + s\,X]\ = \ m + s\,\mathrm{E}[X] \\
    \sigma^2 &\ =\ \mathrm{var}[m + s\,X]\ = \ s^2\,\mathrm{var}[X].
\end{split}
\end{equation}
Hence we have
\begin{equation}\label{eq:solv}
\begin{split}
   s  &\ =\ \frac{\sigma}{\sqrt{\mathrm{var}[X]}},\\
    m &\ =\ \mu - \frac{\sigma}{\sqrt{\mathrm{var}[X]}}\,\mathrm{E}[X].
\end{split}
\end{equation}
\end{subequations}
Quite often in \eqref{eq:solv}, the variable $X$ will be standardised such that it follows that $m = \mu$ and $s = \sigma$. 
In particular, this applies to the case where $X$ is standard normal. Occasionally, it may be inconvenient to 
standardise $X$ as in case of the t-distribution family with fixed degree of freedom $n > 2$. Then the 'standard' 
distribution of the family would be characterised by $\mathrm{E}[X] = 0$ and $\mathrm{var}[X] = \frac n{n-2} > 1$. 

\paragraph{Exponential transformation of location-scale families.} As discussed in Section~\ref{se:ESLocSc},
for generating two-parameter families of distributions
on the positive real half-axis is convenient to take the exponential of a location-scale family in 
the sense of Definition~\ref{de:LocScale}. In general, in this case there is no longer a 
closed-form solution to the mean-variance fitting problem. In the important special case where 
$X$ is a standard normal variable,
\eqref{eq:ExpLocSc} just gives the family of lognormal distributions. 
It is common knowledge that the mean-variance fitting
problem
 \begin{subequations}
\begin{equation}
\begin{split}
    \mu &\ =\ \mathrm{E}[\exp(m + s\,X)]\ = \ \exp(m + s^2/2), \\
    \sigma^2 &\ =\ \mathrm{var}[\exp(m + s\,X)]\ = \ \bigl(\exp(s^2)-1\bigr)\,\exp(2\,m + s^2),
\end{split}
\end{equation}
with $\mu >0$ and $\sigma > 0$ then has the following solution \citep[see, e.g.,][]{wiki:lognormal}:
\begin{equation}\label{eq:ExpSolv}
\begin{split}
   s  &\ =\ \sqrt{\log\left(\frac{\sigma^2}{\mu^2}+1\right)},\\
    m &\ =\ \log\left(\frac{\mu^2}{\mu^2+\sigma^2}\right).
\end{split}
\end{equation}
\end{subequations}

\paragraph{Normal transformation of location-scale families.} 
Two-parameter families of distributions on the unit interval can be generated by applying a mapping of $\mathbb{R}$ onto $(0,1)$
to a location-scale distribution family in the sense of Definition~\ref{de:LocScale}. In principle, each 
distribution function of a distribution on the entire real axis (or link function) would do the job. However, for arbitrary
combinations of link function and location-scale distribution families, in general there are no closed-form solutions
for the mean and the variance. 

The case of the normal location-scale family with standard normal link function (denoted by $\Phi$ in the following) 
is a notable exception. We are then studying distributions of the form 
\begin{equation}\label{eq:Vasicek}
Y \ = \ \Phi(m + s\,X),
\end{equation}
where $X$ denotes a standard normal random variable and we assume $m \in \mathbb{R}$ and $s > 0$. With the alternative 
parametrisation 
$$m\ = \ \frac{\Phi^{-1}(p)}{\sqrt{1-\varrho}}, \quad
s \ = \ \sqrt{\frac{\varrho}{1-\varrho}}, \quad 0<p<1, 0<\varrho < 1,$$
the distribution of $Y$ given by \eqref{eq:Vasicek} is called \emph{Vasicek distribution} by 
some authors. It is easy to show 
\citep[see, e.g.,][]{meyer2009estimation} that 
\begin{equation}\label{eq:VasicekMom}
\begin{split}
   \mathrm{E}[\Phi(m + s\,X)]  &\ =\ \Phi(m),\\
    \mathrm{var}[\Phi(m + s\,X)] &\ =\ \Phi_2\bigl(m, m; \tfrac{s^2}{1+s^2}\bigr) - \Phi(m)^2,
\end{split}
\end{equation}
with $\Phi_2(\cdot, \cdot; r)$ denoting the bivariate standard normal distribution with correlation $r$.
Note that for any random variable $Z$ with values in $[0,1]$ and $\mathrm{P}[0<Z<1] > 0$ it holds that
\begin{equation}\label{eq:restrict}
0 < \mathrm{E}[Z] < 1 \quad\text{and}\quad 0 < \mathrm{var}[Z] < \mathrm{E}[Z]\,(1-\mathrm{E}[Z]).
\end{equation}
As a consequence of \eqref{eq:VasicekMom} and \eqref{eq:restrict}, it can be shown that for any pair $(\mu, \sigma)$ 
with $0 < \mu < 1$ and $0 < \sigma^2 < \mu\,(1-\mu)$ there is exactly one solution $(m, s)$ of the equation
system
\begin{equation}\label{eq:VasicekSys}
\begin{split}
   \mu  &\ =\ \Phi(m),\\
    \sigma^2 &\ =\ \Phi_2\bigl(m, m; \tfrac{s^2}{1+s^2}\bigr) - \Phi(m)^2.
\end{split}
\end{equation}
This solution is given by $m = \Phi^{-1}(\mu)$ and $s = \sqrt{\frac{r}{1-r}}$ where $r$ is the unique solution in $(0,1)$ of the
equation
$$0 \ =\ \Phi_2(m, m; r) - \Phi(m)^2 - \sigma^2.$$
Hence there is no fully explicit solution of the mean-variance fit problem for the Vasicek distribution.
That is why despite the economic credentials of the Vasicek distribution \citep{meyer2009estimation} in practice often
the two-parameter family of the Beta distributions is used for mean-variance fit in the unit interval.

The Beta distribution with parameters $\alpha >0$ and $\beta > 0$ is specified by its density
$$b_{\alpha, \beta}(x) \ = \ \begin{cases}
\frac{\Gamma(\alpha+\beta)}{\Gamma(\alpha)\,\Gamma(\beta)}\, x^{\alpha-1}\,(1-x)^{\beta-1}\,dt, & 
    0 < x < 1,\\
 0, & x \le 0 \ \text{or}\ x \ge 1,
\end{cases}$$
where $\Gamma$ denotes the Gamma-function $\Gamma(a) = \int_0^\infty t^{a-1}\,e^{-t}\,dt$ for $a > 0$.
If $\mu$ and $\sigma$ with $0 < \mu < 1$ and $0 < \sigma^2 < \mu\,(1-\mu)$ are given and $Y$ denotes 
a Beta-distributed random variable, the parameters $\alpha$ and 
$\beta$ of the distribution of $Y$ can be determined by solving the equation system
\begin{subequations}
\begin{equation}
\begin{split}
    \mu &\ =\ \mathrm{E}[Y]\ = \ \frac \alpha{\alpha+\beta}, \\
    \sigma^2 &\ =\ \mathrm{var}[Y]\ = \ \frac{\alpha\,\beta}{(\alpha+\beta)^2\,(\alpha+\beta+1)},
\end{split}
\end{equation}
for $\alpha$ and $\beta$. It turns out \citep[see, e.g.,][]{wiki:beta} that there is a closed-from solution given by
\begin{equation}\label{eq:BetaSolv}
\begin{split}
   \alpha  &\ =\ \mu \left(\frac{\mu\,(1-\mu)}{\sigma^2}-1\right),\\
    \beta &\ =\ (1-\mu) \left(\frac{\mu\,(1-\mu)}{\sigma^2}-1\right).
\end{split}
\end{equation}
\end{subequations}

\end{document}